\documentclass[aps,superscriptaddress,twocolumn]{revtex4}
\usepackage{graphicx,bm,amsmath,amsfonts,amssymb}
\usepackage{epstopdf}
\newcommand{\Kor}{{\rm Kor}}
\newcommand{\Ueh}{U_{\rm eh}}
\newcommand{\barH}{\bar H}

\newcommand{\Uee}{U}
\newcommand{\Jeh}{J_{\rm eh}}
\newcommand{\projected}{{\rm p}}
\newcommand{\electron}{{\rm e}}
\newcommand{\const}{{\rm const.}}
\newcommand{\hole}{{\rm h}}

\newcommand{\SOMINFO}{Appendix}
\newcommand{\SOMDIS}{Appendix}

\newcommand{\initial}{{\rm i}}
\newcommand{\final}{{\rm f}}
\newcommand{\tunnel}{{\rm t}}
\newcommand{\QD}{{\rm QD}}
\newcommand{\ground}{{\rm G}}

\newcommand{\aaaa}{{a}}
\newcommand{\threshold}{{\rm th}}
\newcommand{\regime}{{r}}
\newcommand{\FO}{{\rm FO}}
\newcommand{\LM}{{\rm LM}}
\newcommand{\SC}{{\rm SC}}
\newcommand{\laser}{{\rm L}}
\newcommand{\Tk}{{T_{\rm K}}}

\newcommand{\gate}{{\rm g}}
\newcommand{\FermiR}{{\rm c}}

\newcommand{\Fermi}{{\rm F}}
\newcommand{\pdag}{{\phantom{\dagger}}}
\newcommand{\qphan}{\quad \phantom{.}}
\newcommand{\qqph}{\quad \phantom{.}}
\newcommand{\efinal}{\varepsilon^\final_{\electron \sigma}}
\newcommand{\magnetization}{{m_\electron^\final}} 

\newcommand{\upperlevel}{{\rm upper}}
\newcommand{\lowerlevel}{{\rm lower}}

\newcommand{\Eq}[1]{Eq.~(\ref{#1})}
\newcommand{\Eqs}[1]{Eqs.~(\ref{#1})}

\newcommand{\be}{\begin{equation}}
\newcommand{\ee}{\end{equation}}
\newcommand{\beq}{\begin{eqnarray}}
\newcommand{\eeq}{\end{eqnarray}}
\newcommand{\ba}{\begin{array}}
\newcommand{\ea}{\end{array}}
\newcommand{\bea}{\begin{eqnarray}}
\newcommand{\eea}{\end{eqnarray}}
\newcommand{\ex}[1]{\mbox{e}^{#1}}

\newcommand{\im}[1]{\mbox{Im}\left[#1\right]}

\newcommand{\bra}[1]{| #1 \rangle}

\newcommand{\bma}{\begin{matrix}}
\newcommand{\ema}{\end{matrix}}

\newcommand{\vareps}{\varepsilon}

\newcommand{\eusi}{\hat{e}_\uparrow}
\newcommand{\edsi}{\hat{e}_\downarrow}
\newcommand{\euD}{\hat{e}_\uparrow^\dag}
\newcommand{\edD}{\hat{e}_\downarrow^\dag}

\newcommand{\esigD}{\hat{e}_\sigma^\dag}
\newcommand{\esignD}{\hat{e}_{-\sigma}^\dag}
\newcommand{\Emu}{\hat{e}_\mu}

\newcommand{\EmuD}{\hat{e}_\mu^\dag}

\newcommand{\Enu}{\hat{e}_\nu}

\newcommand{\EnuD}{\hat{e}_\nu^\dag}
\newcommand{\EnunD}{\hat{e}_{-\nu}^\dag}

\newcommand{\hdD}{\hat{h}_\Downarrow^\dag}

\newcommand{\cdk}{\hat{c}_{k,\downarrow}}
\newcommand{\csigk}{\hat{c}_{k,\sigma}}

\newcommand{\cukD}{\hat{c}_{k,\uparrow}^\dag}
\newcommand{\cdkD}{\hat{c}_{k,\downarrow}^\dag}

\newcommand{\csigkD}{\hat{c}_{k,\sigma}^\dag}

\newcommand{\csignq}{\hat{c}_{{k'},-\sigma}}

\newcommand{\csigqD}{\hat{c}_{{k'},\sigma}^\dag}

\newcommand{\cmuQD}{\hat{c}_{q,\mu}^\dag}

\newcommand{\cnunk}{\hat{c}_{k,-\nu}}

\newcommand{\cnuqD}{\hat{c}_{{k'},\nu}^\dag}

\newcommand{\eff}{{\rm eff}}

\newcommand{\GlazmanPustilnik}{\cite{Glazman2005}}
\newcommand{\Gunnarsson}{\cite{Gunnarsson1983}}
\newcommand{\Helmes}{\cite{Helmes2005}}
\newcommand{\Anderson}{\cite{Anderson1967}}
\newcommand{\Hoegele}{\cite{hoegele04}}
\newcommand{\Tureci}{\cite{TureciTI07}}
\newcommand{\Anders}{\cite{Anders2005}}
\newcommand{\Weichselbaum}{\cite{Weichselbaum2007}}
\newcommand{\Krishna}{\cite{KrishnamurthyWW1982b}}
\newcommand{\Nozieres}{\cite{Nozieres1974}}

\newcommand{\Ambrumenil}{\cite{Ambrumenil2005}}
\newcommand{\Hopfield}{\cite{Hopfield1969}}

\newcommand{\Garst}{\cite{Garst2005}}

\newcommand{\eqFGR}{2}
\newcommand{\eqALM}{6}
\newcommand{\eqHopfield}{8}

\begin{document}
\title{Shedding light on non-equilibrium dynamics of a spin coupled to fermionic reservoir}
\author{Hakan E. T\"ureci}
\email{tureci@phys.ethz.ch}
\affiliation{Institute for Quantum Electronics, ETH-Z\"urich, CH-8093 Z\"urich, Switzerland}
\author{M. Hanl}
\affiliation{Arnold Sommerfeld Center for Theoretical Physics, Ludwig-Maximilians-Universit\"at M\"unchen, D-80333 M\"unchen, Germany}
\author{M. Claassen}
\affiliation{}
\author{A. Weichselbaum}
\author{T. Hecht}
\affiliation{Arnold Sommerfeld Center for Theoretical Physics, Ludwig-Maximilians-Universit\"at M\"unchen, D-80333 M\"unchen, Germany}
\author{B. Braunecker}
\affiliation{Department of Physics, University of Basel, Klingelbergstrasse 82, 4056 Basel, Switzerland}
\author{A. Govorov}
\affiliation{Department of Physics and Astronomy, Ohio University, Athens, Ohio 45701, USA}
\author{L. Glazman}
\affiliation{Sloane Physics Laboratory, §Yale University, New Haven, CT 06520, USA}
\author{J. von Delft}
\affiliation{Arnold Sommerfeld Center for Theoretical Physics, Ludwig-Maximilians-Universit\"at M\"unchen, D-80333 M\"unchen, Germany}
\author{A. Imamoglu}
\affiliation{Institute for Quantum Electronics, ETH-Z\"urich, CH-8093 Z\"urich, Switzerland}

\date{\today}
\begin{abstract}
\end{abstract}
\maketitle

{\bf A single confined spin interacting with a solid-state environment has emerged as one of the fundamental paradigms of mesoscopic physics. In contrast to standard quantum optical systems, decoherence that stems from these interactions can in general not be treated using the Born-Markov approximation at low temperatures. Here we study the non-equilibrium dynamics of a single-spin in a semiconductor quantum dot adjacent to a fermionic reservoir and show how the dynamics can be revealed in detail in an optical absorption experiment. We show that the highly asymmetrical optical absorption lineshape of the resulting Kondo exciton consists of three distinct frequency domains, corresponding to short, intermediate and long times after the initial excitation, which are in turn described by the three fixed points of the single-impurity Anderson Hamiltonian. The zero-temperature power-law singularity dominating the lineshape is linked to dynamically generated Kondo correlations in the photo-excited state. We show that this power-law singularity is tunable with gate voltage and magnetic field, and universal.}

Even though quantum dots (QD) are commonly referred to as artificial
atoms, their physical properties can be substantially different from
that of real atoms. In particular, a single electron spin confined in
a QD is subject to hyperfine interactions with the QD nuclear spin
ensemble and exchange interactions with the Fermi gas that controls
the charging state of the QD. For many applications, such as those
aimed at using QD spins to represent quantum information, the nuclear
spin ensemble and the Fermi gas of electrons are treated as
\emph{reservoirs} that induce decoherence of the confined electron
spin. Tremendous progress in understanding and controlling spin
decoherence induced in particular by QD hyperfine interactions has
been achieved\cite{PettaJTLYLMHG2005}. In contrast, relatively few
studies have addressed the real-time dynamics of an optically excited
QD electron coupled to a nearby fermionic reservoir (FR).

The role of coherent tunnel coupling between a QD and a FR has been
investigated since the late 1990s in the context of low-temperature
transport spectroscopy\cite{GoldhaberSMAMK98, CronenwettOK98,
  KouwenhovenG01, Glazman2005} and is known to lead to one of the most
spectacular phenomena of many-body physics -- the Kondo effect\cite{Kondo1964}. The conductance through a QD in the Coulomb
blockaded regime is proportional to the local density of states (LDOS)
of the QD level; the Kondo effect introduces a quasiparticle peak in
the LDOS, the Kondo resonance, at the Fermi energy $\vareps_\Fermi$,
which leads to enhanced linear conductance through the QD.  In
contrast, \emph{optical signatures} of the Kondo effect in quantum
dots have so far not been observed experimentally\cite{Smith2005,Dalgarno2008}. Prior theoretical work on this topic is
also scarce, yet point to novel signatures: Govorov and co-workers
predicted\cite{GovorovKW03}, using a simplified approach based on
variational wave functions\cite{Gunnarsson1983}, that
doubly charged QDs could exhibit optical resonances whose width
depends on an energy scale that is determined by the exchange
interactions. Helmes \emph{et al.}~used Wilson's numerical
renormalization group (NRG) method to calculate the absorption and
emission lineshapes of excitons in the presence of Kondo correlations
(henceforth called Kondo excitons), finding power-law divergences near
threshold\cite{Helmes2005}.

Here we analyze the non-equilibrium dynamics following a quantum quench of a QD spin that is coupled to a FR. In a quantum quench, parameters of the Hamiltonian of an interacting quantum system are changed over a very short time scale. We show that an optical absorption experiment naturally implements such a quantum quench, and the measured optical absorption lineshape reveals, in a uniquely direct way, how for intermediate and long timescales standard quantum optical techniques such as Markov and Born approximations fail to describe the dynamics.
Since the absorption of a photon \emph{suddenly} switches on an attractive potential that (under suitable conditions) favors Kondo correlations and spin screening, the dynamic emergence of such correlations with time can be probed in ways not possible in transport experiments. For short times (large detunings), charge fluctuations dominate and coupling to the FR can be described using the Born-Markov approximation\cite{foot1}. For intermediate times (intermediate detunings) spin fluctuations dominate; since the spin excitations of the FR are long-lived, the dynamics in this regime is non-Markovian. For long times (small detunings), the orthogonality catastrophe physics\cite{Anderson1967} dominates and leads to a complete breakdown of the Born-Markov approximation.  Remarkably, the resulting highly asymmetrical and singular absorption lineshape directly maps out these three regimes, which we link to the three fixed points of the single-impurity Anderson Hamiltonian. We find that the power-law singularity in the lineshape is tunable with gate voltage and magnetic field, and universal.

\begin{figure}
\includegraphics[clip,width=0.9\linewidth]{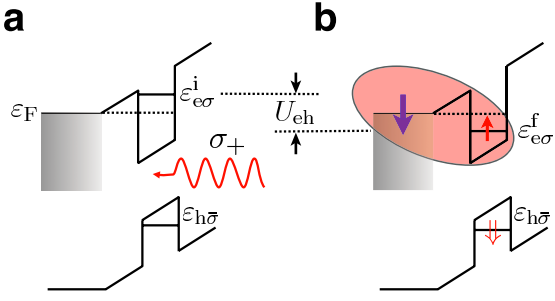}
\caption{{\bf Schematics of the absorption configuration}. 
The fermionic
  reservoir in the back-contact (of e.g.\ a Schottky-diode structure)
  couples via charge tunneling to the localized electronic degrees of
  freedom of the quantum dot (see \SOMDIS~1). Starting from {\bf a} an
  uncharged QD at $t=0$, optical excitation to a {\bf b} neutral exciton
  state ($X^{0}$) at $t=0^{+}$ featuring a single electron in the e-level of the QD (along with a hole in the h-level) which may develop Kondo correlations with the fermionic reservoir in the long-time limit.}
\label{fig:setup}
\end{figure}

\paragraph*{\bf Model for a tunable, optically active quantum impurity -- the excitonic Anderson Model.} 

We consider a QD, tunnel-coupled to a FR (see Fig.~1), whose charge state is controllable via an external gate
voltage $V_\gate$ applied between a top Schottky gate and the
FR (see \SOMDIS~1 for details). Assume $V_\gate$ to be tuned such that optical absorption occurs via the so-called neutral exciton transition\cite{hoegele04} ($X^{0}$),
in which an electron-hole pair is created in the localized
$s$-orbitals of the QD's conduction- and valence bands (to be called
e- and h-levels, respectively). For circularly polarized light with
frequency $\omega_\laser$ propagating along the $z$-axis of the
heterostructure\cite{TureciTI07}, the QD-light interaction is
described by $H_\laser \propto (e^\dagger_\sigma h^\dagger_{\bar
  \sigma} \ex{-i \omega_\laser t} + {\rm h.c.})$, where $e^\dagger_\sigma/h^\dagger_{\bar
  \sigma}$ create an electron/hole in the e- and h-levels respectively and $\sigma = -
\bar \sigma \in \{+,-\}$. We model the
system before/after absorption by the initial/final Hamiltonian
$H^{\initial/ \final} = H^{\initial / \final}_\QD + H_\FermiR +
H_\tunnel$, where
\begin{equation}
\label{eq:Hinitial-final}
H_\QD^\aaaa = \sum_\sigma \varepsilon^\aaaa_{\electron
  \sigma} n_{\electron \sigma} + \Uee n_{\electron \uparrow}
n_{\electron \downarrow} + \delta_{\aaaa \final}
\varepsilon_{\hole \bar \sigma} \;  \quad (a = \initial, \final)
\end{equation}
describes the QD, with Coulomb cost $\Uee$ for double occupancy of the
e-level, $n_{\electron \sigma} = e_\sigma^\dagger e_\sigma^\pdag$, and
hole energy $\varepsilon_{\hole \bar \sigma}$ ($ > 0$, on the order of the
band gap). The e-level's initial and final energies before and after
absorption, $\varepsilon^{\aaaa}_{\electron \sigma}$ ($\aaaa =
\initial, \final$), differ by the Coulomb attraction $\Ueh (>0)$
between the newly created electron-hole pair, which pulls the final
e-level downward, $\varepsilon^\aaaa_{\electron \sigma} =
\varepsilon_{\electron \sigma} - \delta_{\aaaa \final} \Ueh$
(Fig.~1b).  This stabilizes the excited electron against decay into
the FR, provided that $\efinal $ lies below the FR's Fermi energy
$\varepsilon_\Fermi = 0 $. $H_\FermiR = \sum_{k \sigma}
\varepsilon_{k \sigma} c^\dagger_{k \sigma} c^\pdag_{k \sigma}$
represents a noninteracting conduction band (the FR) with half-width
$D= 1/(2 \rho)$ and constant density of states $\rho \,
(\varepsilon_k) = \rho \theta (D - |\varepsilon_k|)$ per spin, while
$H_\tunnel = \sqrt{\Gamma/\pi \rho} \sum_\sigma (e^\dagger_{\sigma} c^\pdag_\sigma
+ {\rm h.c.})$, with $c_\sigma = \sum_k c_{k \sigma}$, describes its
tunnel-coupling to the e-level, giving it a width $\Gamma$. A
magnetic field $B$ along the growth-direction of the heterostructure
(Faraday configuration) causes a Zeeman splitting, 
$\varepsilon_{\electron \sigma} = \varepsilon_\electron + \frac{1}{2}
\sigma g_\electron B$, $\varepsilon_{\hole \sigma} = \varepsilon_\hole
+ \frac{3}{2} \sigma g_\hole B$ (see \SOMDIS~1; the Zeeman splitting of FR
states can be neglected for our purposes, see \SOMDIS~8.)
We set $\mu_{\rm B} = \hbar = k_{\rm B} = 1$, give energies
in units of $D=1$ throughout, and assume $T, B \ll \Gamma \ll \Uee,
\Ueh \ll D \ll \varepsilon_{\hole \bar \sigma}$. A realistic set of
parameters would be, e.g.\ $\Gamma \approx 1$-$10$ meV, $\Uee \approx \Ueh 
\approx 15$-$25$ meV (in general $\Ueh$ is slightly larger than $\Uee$,
resulting in a ``trionic redshift'' of the order of $\approx 5$ meV), $D
\approx 30$ meV, $\varepsilon_{\hole \bar \sigma} \approx 1.3$ eV, 
$g_\electron \approx - 0.6$-0.7,
$g_\hole \approx 1.1$-1.2.

We focus on the case where the e-level is essentially empty in the
initial state and singly-occupied in the ground state of the final
 Hamiltonian, $\bar n^\initial_\electron \simeq 0$ and $\bar
n^\final_\electron \simeq 1$.  (Here $\bar n^\aaaa_\electron = \langle
n_\electron \rangle_\aaaa$ is the thermal average with respect to
$H^\aaaa$ of $n_\electron = \sum_\sigma n_{\electron \sigma}$.)  This
requires $\varepsilon_{\electron \sigma}^\initial \gg \Gamma$, and $-
\Uee + \Gamma \lesssim \efinal \lesssim - \Gamma$. The initial ground
state, needed below, will thus be approximated by the free Fermi sea,
$|\ground \rangle_\initial \simeq \prod_{\varepsilon_{k\sigma} <
  \varepsilon_\Fermi} c_{k \sigma}^\dagger |{\rm Vac} \rangle$,
neglecting terms of order
$\Gamma/\varepsilon^\initial_{\electron\sigma}$. In particular, some
(but not all) parts of the text will focus on the case that $H^\final$
represents the \emph{symmetric} excitonic Anderson model
($H^\final$=SEAM), with $\efinal = - \Uee/2$, for which $\bar
n_\electron^\final = 1$ exactly.

\begin{figure}
\includegraphics[clip,width=\linewidth]{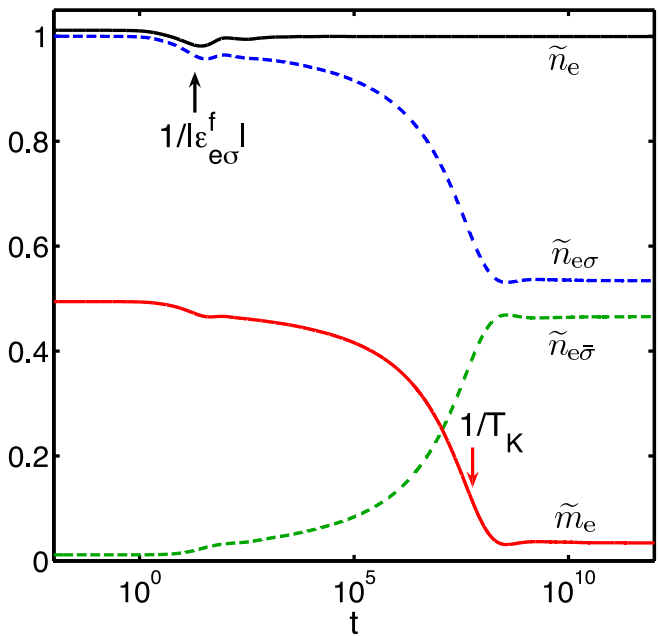}
\caption{{\bf Non-equilibrium time evolution of spin and charge of the photo-excited electron}. Plotted is the non-equilibrium
  time-evolution of charge and spin degrees of freedom of the
  photo-excited electron after the sudden creation of an $e^\dagger_+
  h^\dagger_-$ exciton at time $t=0$. While fluctuations in both e-level's total charge $\widetilde n_\electron$ and spin set in around the time scale $1/|\varepsilon^\final_\sigma|$, the equilibration of the
  spin-$\sigma$ populations $\widetilde n_{\electron \sigma}(t)$ and
  screening of the local spin $\widetilde m_\electron (t)$ sets in on
  the time scale $1/\Tk$.  The deviations (by about 3\%) of
  $\widetilde n_{\electron} (\infty)$, $\widetilde n_{\electron\sigma}
  (\infty)$ and $\widetilde m_\electron (\infty)$ from their expected
  equilibrium values (1, $\frac{1}{2}$ and 0, respectively, for the
  case $H^\final$=SEAM depicted here), are known artifacts of
  time-dependent NRG\cite{Anders2005}, presumably due to the NRG
  discretization scheme, which is inevitably coarse at large energies
  (see \SOMDIS~3).  Here, $t$ is measured in units of $1/D$. NRG parameters: $\Uee=0.1D$, $\varepsilon^\initial_{\electron} = 0.75 \Uee$, $\varepsilon^\final_{\electron} = -0.5 \Uee$, $\Gamma=0.03 \Uee$, $\Tk = 5.9 \cdot 10^{-6} \, \Gamma$, $T=0$, $B=0$, $\Lambda = 1.8$, Kept states: $1024$, $\alpha = 0.4$.}
\label{fig:t-NRG}
\end{figure}

\paragraph*{\bf Time evolution of the charge and spin after a quantum quench induced by absorption.} To gain intuition for how the system would
respond to the \emph{sudden} creation of an $e^\dagger_\sigma
h^\dagger_{\bar \sigma}$ exciton at time $t=0$, it is instructive to
calculate the subsequent time evolution of the average charge
$\widetilde n_\electron (t) = (\widetilde n_{\electron +} + \widetilde
n_{\electron - } )(t)$ and spin $\widetilde m_\electron (t) =
\frac{1}{2} (\widetilde n_{\electron +} - \widetilde n_{\electron -})
(t)$ of the e-level, where $\widetilde n_{\electron \sigma'} (t) =
{\rm Tr} ( e^{-i H^\final t} \hat \rho^\final_\projected e^{i H^\final
  t} n_{\electron \sigma'})$ and $\hat \rho^\final_\projected =
e^\dagger_\sigma \hat \rho^\initial e_\sigma / (1 - \widetilde
n^\initial_{\electron \sigma})$ is a projected version of the initial
density matrix, normalized such that $\widetilde n_{\electron \sigma}
(0) = 1$.  Fig.~2 shows a typical result for $T=0$ and
$H^\final$=SEAM, obtained using time-dependent NRG\cite{Anders2005}
(see Methods and \SOMDIS~3).  The non-equilibrium dynamics following such a quantum quench
shows two distinct time scales: (i) Fluctuations in both charge and spin set in around
the time scale $t \simeq 1/|\efinal|$ associated with virtual
transitions of electrons between e-level and FR.  Whereas the charge
equilibrates (towards 1) shortly thereafter, (ii) the spin decays
(towards $\simeq 0$) much more slowly, on the scale $t \simeq 1/\Tk$,
where $\Tk = \sqrt{\Gamma U/2} e^{- \pi |\varepsilon^\final_\electron
  (\varepsilon^\final_\electron + \Uee)|/(2 U \Gamma)}$ is the Kondo
temperature\cite{Schrieffer1966,Haldane1978} associated with $H^\final$. (For finite temperatures, the time
scale on which $\widetilde m_\electron(t)$ decays is $\min\{1/\gamma_\Kor
,1/\Tk \}$, where $\gamma_\Kor = T/ \ln^2 (T/\Tk)$ is the Korringa
relaxation rate\cite{Glazman2005}.)  The decay is due to spin-flip
processes, mediated by electrons of opposite spin hopping between
e-level and FR, leading to non-Markovian dynamics because the bath
remembers its spin state between two e-level spin-flips. As a result a
FR screening cloud builds up over time, which ultimately screens the
localized spin into a singlet. Note that this in particular means that
for $t>1/T_{K}$ the ``reservoir'' (FR) is substantially modified,
implying the complete breakdown of the Born-Markov approximation.  We will come
back to this point further below.

The time-evolution depicted in Fig.~2 could in principle be observed
by a $\pi$-pulse excitation of the QD followed by polarization
resolved detection of the photoluminescence.  However, the
fingerprints of the non-equilibrium dynamics can be more clearly
discerned by measuring the absorption lineshape of a continuous-wave
laser field, as we show next.

\paragraph*{\bf Absorption lineshape of a Kondo exciton.} Absorption sets in once
$\omega_\laser$ exceeds a threshold frequency
$\omega_\threshold = E^\final_\ground - E^\initial_\ground$, which is on the order of $\efinal
 + \varepsilon_{\hole \bar \sigma}$ (minus corrections
due to tunneling and correlations).  By Fermi's
golden rule the absorption lineshape at temperature $T$
and detuning $\nu = \omega_\laser - \omega_\threshold$ is proportional
to
\begin{eqnarray}
A_\sigma (\nu) &= &  2\pi \sum_{mn} \rho_m^\initial
\left| {}_\final \langle n | e^\dagger_\sigma
| m \rangle_\initial \right|^2
\delta (\omega_\laser - E_n^\final + E_m^\initial ) .
\qphan
\label{eqabslsexact}
\end{eqnarray}
Here $|m\rangle_\aaaa$ and $E_m^\aaaa$ are the exact eigenstates and
-energies of $H^\aaaa$ and $\rho^\initial_m = \ex{-E^\initial_m/T} /
Z^\initial$ the initial Boltzmann weights.  For future reference, we
note that \Eq{eqabslsexact} can be expressed as $A_\sigma (\nu) = - 2
{\rm Im} {\cal G}_{\electron \electron}^\sigma (\nu)$, where, for
$T=0$,
\begin{eqnarray}
\label{eq:T=0Gee}
{\cal G}_{\electron \electron}^\sigma (\nu) =
  {}_\initial \langle \ground | e^\pdag_\sigma
  \frac{1}{\nu_+ - \barH^\final}
 e^\dagger_\sigma | \ground \rangle_\initial \; ,
\end{eqnarray}
with $\nu_+ = \nu + i 0$ and $\barH^\final = H^\final -
E^\initial_\ground - \omega_\threshold$. Moreover, the 
Fourier representation ${\cal G}_{\electron \electron}^\sigma (\nu) =
\int {\rm d} t \, e^{i t(\nu_+ + \omega_\threshold)} G_{\electron
  \electron}^\sigma (t)$, where 
$ G_{\electron  \electron}^\sigma (t) 
= - i \theta (t) {}_\initial \langle \ground |
e^{i H^\initial t} e_\sigma e^{-i H^\final t} e^\dagger_\sigma |
\ground \rangle_\initial$, makes explicit that this correlator directly probes
the dynamics, described above and in Fig.~2, of a photo-generated
electron coupled to a FR (see {\SOMINFO} for details).

\begin{figure*}
\includegraphics[clip,width=\linewidth]{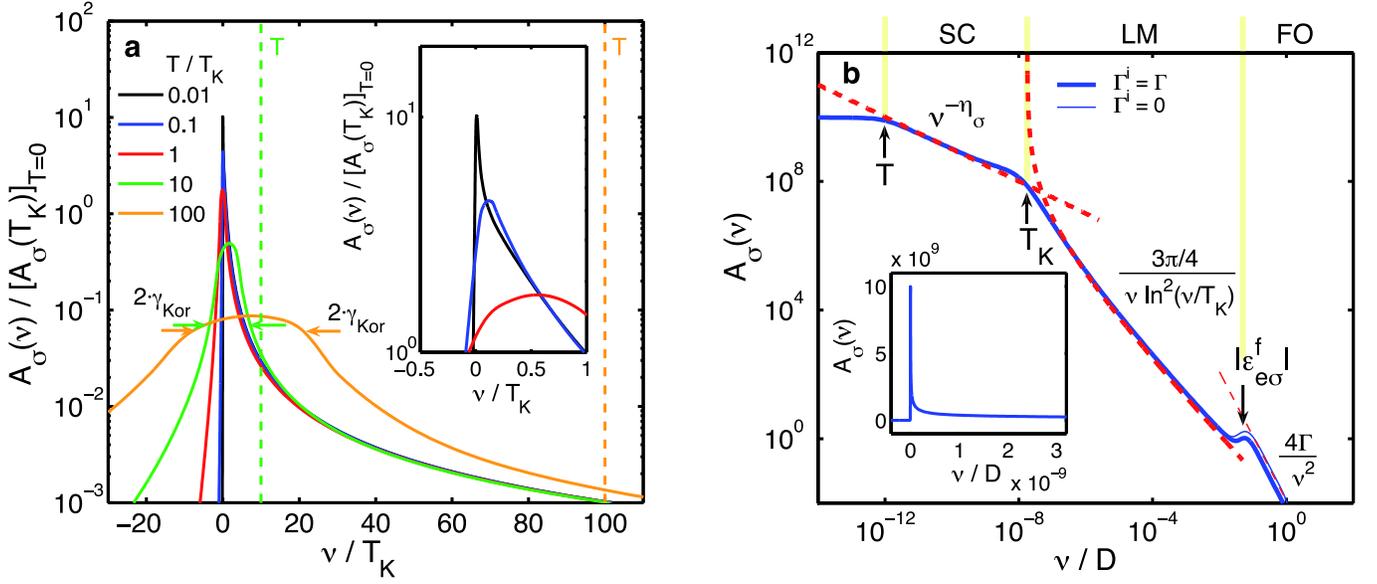}
\caption{{\bf NRG vs. analytic results for the $B=0$ absorption
  lineshape}. Solid lines represent the absorption lineshapes for
  $H^\final$=SEAM calculated by NRG {\bf a} for several temperatures on a
  semi-log plot (See \SOMDIS~5), and {\bf b} at very low temperature
  (blue) on a log-log plot (inset shows the same lineshape in a linear
  plot). The latter reveals three distinct functional forms (red dashed
  lines) for high, intermediate and small detuning, labeled \FO, \LM\
  and \SC, respectively as captured by the fixed point perturbation
  theory (Eqs.~(\ref{eq:A-FO-results}), (\ref{eq:A-LM-results}) and
  (\ref{eq:ASC})) (these, and thin blue line, were calculated assuming $\Gamma^\initial = 0$). Arrows and light yellow
  lines indicate the crossover scales $T$, $\Tk$ and $|\efinal|$. NRG parameters: $\Uee=0.1D$, $\varepsilon^\initial_{\electron} = 0.75U$, $\varepsilon^\final_{\electron} = -0.5 \Uee$, $\Gamma=0.03 \Uee$, $\Tk = 5.9 \cdot 10^{-6} \, \Gamma$, $T= 3.3 \cdot 10^{-10} \, \Gamma$ for {\bf b}, $B=0$, $\Lambda = 1.8$, Kept states: $1024$, $\alpha = 0.4$ for {\bf a} and $\alpha = 0.5$ for {\bf b}. }
\label{fig:absNRG0}
\end{figure*}

We used NRG to calculate $A_\sigma (\nu)$ from \Eq{eqabslsexact},
generalizing the approach of Ref.~\cite{Helmes2005} to $T\neq 0$ by
following Ref.~\cite{Weichselbaum2007} (see Methods and \SOMDIS~2 for
details). For clarity, we focus first on $H^\final$=SEAM with
$B=0$. Fig.~3a shows a typical result: As temperature is gradually
reduced, an otherwise symmetric lineshape develops into a highly
asymmetric one, dramatically increasing in peak-height as $T
\rightarrow 0$. At $T=0$, the lineshape displays a threshold behavior,
vanishing for $\nu < 0$ and {\em diverging} as $\nu$ tends to 0 from
above. Fig.~3b analyzes this divergence on a log-log plot, for the
case that $T$, which cuts off the divergence, is smaller than all
other relevant energy scales (hence $T=0$ in all analytical
calculations below). Three distinct functional forms are discernible
in the regimes of ``large'', ``intermediate'' or ``small'' detuning,
labeled (for reasons discussed below) \FO, \LM\ and \SC, respectively:
\begin{subequations}
  \label{subeq:lineshape-results}
  \begin{eqnarray}
  \label{eq:large-detuning}
 {\rm (\FO)} \;
& | \efinal |
\lesssim \nu \lesssim D
&: \;  A \propto \nu^{-2}
\theta (\nu - |\efinal|) \; ;
\\
  \label{eq:intermediate-detuning}
 {\rm (\LM)} \;
& \Tk \lesssim \nu \lesssim |\efinal |
&: \; A \propto
\nu^{-1} \ln^{-2} (\nu/\Tk) ; \; \qphan
\\
\label{eq:small-detuning}
 {\rm (\SC)} \;
& T \lesssim \nu \lesssim \Tk
&:  \; A \propto
\nu^{-\eta_\sigma} \; .
  \end{eqnarray}
\end{subequations}

A central goal of this paper is to explain the remarkable series of
cross-overs described above. To this end we note that absorption at
large, intermediate or small detuning probes excitations at
successively smaller energy scales, corresponding to ever longer time
scales after absorption, for which $H^\final$ can be represented by
expansions $ H^\ast_\regime + H^\prime_\regime$ around the three
well-known fixed points\cite{KrishnamurthyWW1982b} of the AM: the
\emph{free orbital}, \emph{local moment} and \emph{strong-coupling}
fixed points ($\regime = \FO, \LM, \SC$), characterized by charge
fluctuations, spin fluctuations and spin screening, respectively.

\paragraph*{\bf Large and intermediate detuning dependence of the lineshape -- perturbative regime.} For large detuning, probing
the time interval $t \lesssim 1/|\efinal|$
immediately after absorption, the \electron-level appears as a free,
filled orbital perturbed by charge fluctuations, described by the
fixed point Hamiltonian $H_\FO^\ast = H_\FermiR + H^\final_\QD $ and
the relevant perturbation $H^\prime_\FO = H_\tunnel$.  Intermediate
detuning probes the times $1/|\efinal|
\lesssim t \lesssim 1/\Tk$ for which real charge fluctuations have
frozen out, resulting in a stable local moment (assuming $\bar
n^\final_\electron \simeq 1$); however, virtual charge fluctuations
still cause the local moment to undergo spin fluctuations, which are
not yet screened. This is described by\cite{Schrieffer1966,KrishnamurthyWW1982b} $H^\ast_\LM = H_\FermiR + \const $ and the RG-relevant
perturbation $H^\prime_\LM = \frac{J}{\rho} \vec s_\electron \cdot \vec
s_\FermiR$ (a potential scattering term in $H^\prime_\LM$, being
RG-irrelevant, will be neglected in the discussion of the
intermediate-detuning regime).
Here $\vec s_j = \frac{1}{2} \sum_{\sigma \sigma'} j^\dagger_\sigma
\vec \tau_{\sigma \sigma'} j^\pdag_{\sigma '} $ (for $j = e,c$), are
spin-operators for the e-level and conduction band, respectively
($\vec \tau$ are Pauli matrices), and $J = 2 \Uee \Gamma / |\pi
\varepsilon_\electron^\final (\varepsilon^\final_\electron + \Uee)|$
is an effective dimensionless exchange constant\cite{Haldane1978}.
Constant contributions to $H^\ast_\regime$ will not be specified,
since they affect only $\omega_\threshold$, whose precise value is not
of present interest.  It suffices to note that for both $\regime =
\FO$ and $\LM$, $e^\dagger_\sigma |\ground \rangle_\initial$ is an
eigenstate of $H^\ast_\regime$ with eigenvalue $\omega_\threshold +
E_\ground^\initial$ (within the accuracy of $H^\ast_\regime$).

For $r = \FO$ and $\LM$, $A_\sigma (\nu)$ can be calculated using
fixed point perturbation theory in $H'_r$, as outlined in Methods and \SOMDIS~5. At
$T=0$, we find
\begin{eqnarray}
\label{eq:A-FO-results}
A^\FO_\sigma (\nu) & =  &
\frac{ 4 \Gamma}{ \nu^2}  \theta(\nu - |\efinal|)
\qquad \; \;
(|\efinal| \lesssim \nu \lesssim D ) ,
\rule[-3mm]{0mm}{0mm} \qphan
\\
\label{eq:A-LM-results}
A^\LM_\sigma (\nu) & =  &
  \frac{3 \pi}{4}
\frac{J^2 (\nu) }{ \nu} \qquad \qquad
\quad
(\Tk \lesssim \nu \lesssim |\efinal| ) . \qphan
\end{eqnarray}
The crossover from $\LM$ to $\FO$ shows up in our numerical results as a small shoulder or side peak in
$A_\sigma (\nu)$ at $|\efinal|$, depending on the detailed choice of parameters (easily seen from $A^\LM_\sigma (|\efinal|) \neq A^\FO_\sigma (|\efinal|)$). This cross-over reflects the fact that for $\nu > |\efinal|$, transitions
from the h-level into unfilled states of the FR are possible, mediated
by $H^\prime_\FO$ and using the e-level (Lorentzian-broadened by
charge fluctuations, hence the $\nu^{-2}$ dependence) as intermediate
state, without creating additional particle-hole excitations.  For
intermediate detuning (\LM) this is not possible; instead, absorption
into the \electron-level is accompanied by the $H_\LM^\prime$-induced
creation of electron-hole pairs in the FR, yielding an additional
phase space factor $\propto \nu$. In \Eq{eq:A-LM-results}, we evoked
scaling arguments\cite{Haldane1978} to replace the bare exchange
constant $J$ by its renormalized, scale-dependent version, $J(\nu) =
\ln^{-1}(\nu/\Tk)$. \Eqs{eq:A-FO-results} and (\ref{eq:A-LM-results})
reproduce \Eqs{eq:large-detuning} and
(\ref{eq:intermediate-detuning}), and quantitatively agree with the
NRG results of Fig.~3.

\begin{figure}
\includegraphics[clip,width=\linewidth]{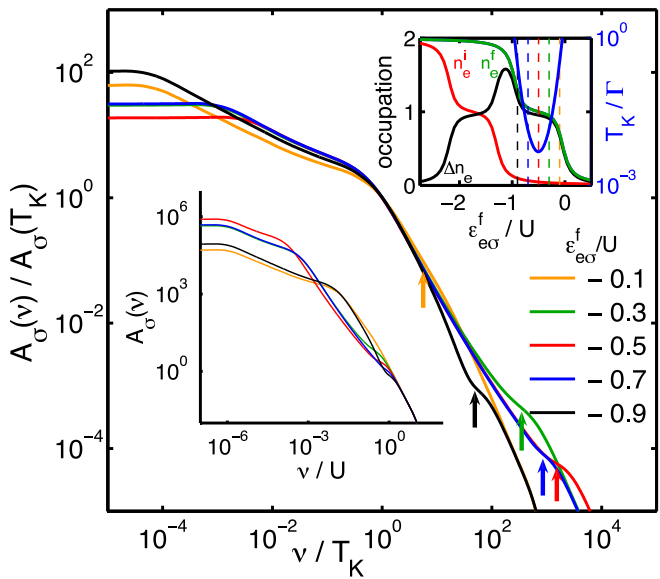}
\caption{{\bf
Universality in the LM-regime}. Lower left
panel: Five lineshapes, corresponding to five different choices of
$\efinal$, indicated by color-coded dashed lines in upper inset, and
arrows in main panel. The inset also gives $\bar
n^\initial_\electron$, $\bar n^\final_\electron$, $\Delta n_\electron$
and $\Tk$ as function of $\efinal$. Main panel: When appropriately
rescaled, the lineshapes collapse onto a universal curve in the
LM-regime $\Tk \lesssim \nu \lesssim |\efinal|$. In the SC-regime $T
\lesssim \nu \lesssim \Tk$, the curves do not collapse, since their
exponents $\eta_\sigma$ depend, via $\Delta n_\electron$, on
$\efinal$. NRG parameters: $\Uee=0.1D$, $\Ueh=1.25 \Uee$, $\Gamma=0.062 \Uee$, $T = 1.6 \cdot 10^{-5} \, \Gamma$, $B=0$, $\Lambda = 1.8$, Kept states: $1024$, $\alpha = 0.5$.}
\label{fig:scalingcollapse}
\end{figure}

For clarity, the above discussion was confined to
$H^\final$=SEAM. However, it can be generalized straightforwardly to
the non-symmetric case with $\efinal \neq - \frac{1}{2} \Uee$, as long as $H^\final$ remains in the LM-regime, with $ \bar
n_\electron^\final \simeq 1$  (see \SOMDIS~4). In the \LM\ regime, the lineshape depends on $\efinal$ and $\Uee$ only
through their influence on $\Tk$, and hence $A^\LM_\sigma (\nu) $ is a
\emph{universal} function of $\nu$ and $\Tk$. This is illustrated in
Fig.~4 for five lineshapes, shown in the lower left panel,
corresponding to different choices of $\efinal$ and hence different
$\Tk$-values (as indicated in inset). When these lineshapes are
rescaled as $A_\sigma (\nu) / A_\sigma (\Tk)$ vs.  $\nu / \Tk$ (main
panel), they collapse onto a universal scaling curve within the LM
regime $\Tk \lesssim \nu \lesssim |\efinal|$.  An experimental
observation of such a scaling collapse would be a smoking gun for the
existence of Kondo correlations.

\paragraph*{\bf Small detuning dependence and the tunable Kondo-edge singularity -- the non-perturbative regime.} As $\nu$ is lowered through the bottom of the
LM regime, $J(\nu)$ increases through unity into the strong coupling
regime, and $A_\sigma (\nu)$ monotonically crosses over to SC
behavior. In this limit, the assumption of a FR that remains unaltered
by the interactions with the QD breaks down, requiring an alternative
approach for determining $A_\sigma (\nu)$.

For small detuning, i.e.\ long times $t > 1/\Tk$, a screening cloud
builds up that tends to screen the local moment into a spin singlet,
as visible in the decay of $\widetilde m_\electron (t \rightarrow \infty)
\rightarrow 0$ (see Fig.~2). The screened spin singlet acts as a
source of strong potential scattering for other FR electrons, causing
the phase of each mode $k\sigma$ to shift by $\delta_{\sigma}
(\varepsilon_{k\sigma})$ relative to its value for $H^\initial$.
This regime can be described by a strong-coupling fixed-point
Hamiltonian $H^\ast_\SC + H^\prime_\SC$ due to Nozi\`eres (given in
\SOMDIS~6). It is formulated purely in terms of these
phase-shifted \FermiR-electrons and makes no reference to
$\electron$-level operators at all, since in the \SC-regime the local
moment is fully screened.  Thus, the fixed-point perturbation strategy
used above cannot be applied here.

This hurdle can be overcome by working in the time-domain, and
relating the correlator $G_{\electron \electron}^\sigma (t)$ mentioned
after \Eq{eq:T=0Gee} to the X-ray edge problem. The latter deals with
the absorption lineshape for an incident X-ray to suddenly dislodge an
electron from an atomic core level, placing it in the conduction band
and leaving behind a core hole (i.e.\ a local scattering
potential). Our situation is similar, in that the screened singlet
also serves as local scattering potential, but more complex, in that
this potential is not turned on suddenly, but emerges only as the
\emph{long-time limit} of the dynamical build-up of a Kondo cloud to
screen the local spin.  Nevertheless, in the long-time limit an
equation of motion approach can be used to relate the correlator
$G_{\electron \electron}^\sigma (t)$ to a similar one, $G_{\FermiR \FermiR}^\sigma (t)$, which involves only conduction band electrons
and whose form is known from the X-ray edge problem (see Methods and \SOMDIS~6
for details).  This readily leads to a power-law divergence
characteristic of X-ray edge problems\cite{Mahan1967,Nozieres1969,Ambrumenil2005},
\begin{equation}
\label{eq:ASC}
A^\SC_\sigma (\nu) \propto T_{\rm K}^{-1} (\nu / \Tk)^{- \eta_\sigma} \; ,
\end{equation}
where the infrared singularity exponent\cite{foot2},
\begin{equation}
\label{eq:exponentsmalldetuning}
\eta_\sigma =
2 \Delta n_{\electron \sigma} - \sum_{\sigma'} (\Delta n_{\electron \sigma'})^2
 \; ,
\end{equation}
depends on the e-level's change in average occupation, $\Delta
n_{\electron \sigma} = \bar n^\final_{\electron \sigma} - \bar
n^\initial_{\electron \sigma}$.  \Eq{eq:exponentsmalldetuning} may be
regarded as a generalized version of ``Hopfield's rule of thumb''\cite{Hopfield1969}, which was established rigorously in Refs.~\cite{Nozieres1969,Affleck1994}. It has an instructive physical
interpretation, based on rewriting it as $\eta_\sigma = 1 -
\sum_{\sigma'} (\Delta n'_{\electron \sigma'})^2$, where $\Delta
n'_{\electron \sigma'} = \Delta n_{\electron \sigma} - \delta_{\sigma
  \sigma'}$ is the charge difference in level e$\sigma'$ between the
final ground state $|\infty \rangle$ (at time $t \to \infty$) and the
state $|0^+ \rangle$ the system finds itself in just after
photo-excitation of a spin-$\sigma$ electron (at $t = 0^+$).  The "1"
in $\eta_\sigma$ represents a $\nu^{-1}$ power law divergence: it may
be thought of as arising from a detuned, virtual transition into a
narrow e-level situated at $\nu = 0$ (giving a Lorentzian detuning
factor $1/\nu^2$), followed by the creation of particle-hole pairs
(with phase space $\nu$) to carry off the excess energy $\nu$,
resulting in a lineshape scaling as $\nu / \nu^2 = \nu^{-1}$. The
$\sum_{\sigma'}(\Delta n'_{\electron \sigma'})^2$ contribution to
$\eta_\sigma$ reflects Anderson orthogonality\cite{Anderson1967}:
since $|\infty \rangle$ and $|0^+\rangle$ have localized
e$\sigma'$-charges that differ by $\Delta n'_{\electron \sigma'}$,
their Fermi reservoir electrons see different scattering potentials,
implying\cite{Anderson1967} that their overlap scales with effective
system size $L \sim \nu^{-1}$ as $\langle \infty | 0^+ \rangle \sim
L^{-\sum_{\sigma'} (\Delta n_{\electron \sigma'})^2}$.
%

\begin{figure*}
\includegraphics[clip,width=\linewidth]{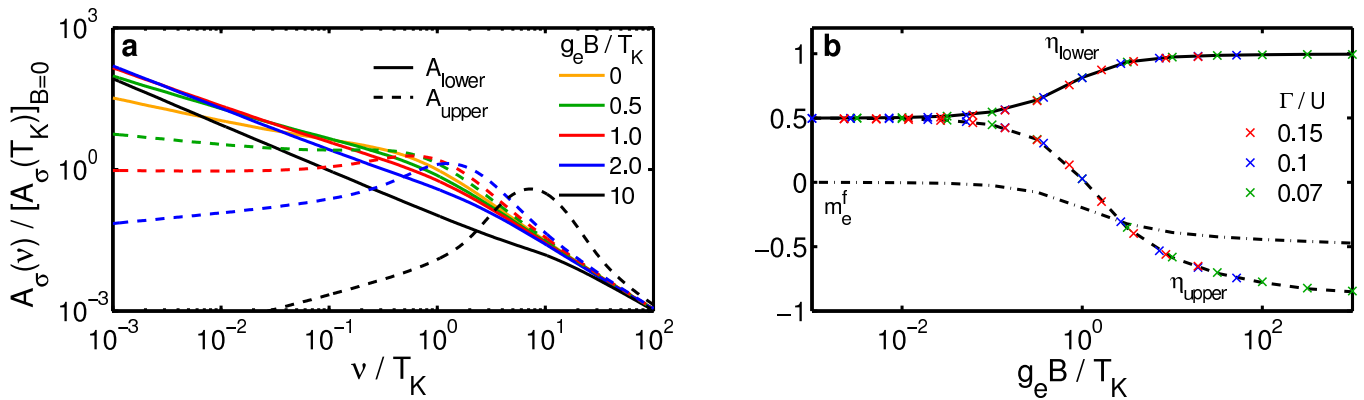}
  \caption{{\bf Magnetic-field dependence of lineshape}.
{\bf a}, Depending on whether the electron excited into the lower or upper Zeeman-split
e-level ($\sigma = \lowerlevel/ \upperlevel$, solid/dashed lines) the
near-threshold divergence, $A_\sigma (\nu) \propto
\nu^{-\eta_\sigma}$, is either strengthened, or suppressed via the
appearance of a peak at $\nu \simeq |g_\electron B|$, respectively.
{\bf b}, Universal dependence on $g_\electron B/\Tk$ of the local moment
$m^\final_\electron$, calculated using the same values as in {\bf a} for
$U$ and $\varepsilon^\final_\electron$ (dash-dotted line), and
the corresponding infrared exponents $ \eta_\lowerlevel$ (solid) and
$\eta_\upperlevel$ (dashed) predicted by
\Eq{eq:exponentsmalldetuning}. Symbols: $\eta_\sigma$-values extracted
from the near-threshold $\nu^{-\eta_\sigma}$ divergence of $A_\sigma
(\nu)$, for several magnetic fields (same color code as in (a)) and
three values of $\Gamma$ (distinguished by symbols).  Symbols and
lines agree to within 1~\%, confirming the applicability of Hopfield's
rule. NRG parameters: $\Uee=0.1D$, $\varepsilon^\initial_{\electron} = 0.75U$, $\varepsilon^\final_{\electron} = -0.5 \Uee$, $\Gamma=0.062 \Uee$, $\Tk = 3.7 \cdot 10^{-3} \, \Gamma$, $T= 0$, $\Lambda = 2.3$, Kept states: $1200$, $\alpha = 0.6$.}
\label{fig:Hopfield}
\end{figure*}

For the X-ray edge problem, $\Delta n_{\electron \sigma}$ and
hence $\eta_\sigma$ are fixed by material parameters (the strength of
the core-hole potential). In contrast, in the present case they can be
\emph{tuned} experimentally by sweeping
$\varepsilon^\initial_{\electron \sigma}$ and $\efinal$ with a gate
voltage or magnetic field (see Figs.~5b and Figure~S3).  This tunability can
be exploited to study universal aspects of Anderson orthogonality
physics that had hitherto been inaccessible.  In particular, if the
system is tuned such that $\bar n^\initial_\electron = 0$ and $\bar
n^\final_\electron = 1$, \Eq{eq:exponentsmalldetuning} can be
expressed as
$\eta_\sigma  =  \frac{1}{2} + 2 \magnetization \sigma
- 2 (\magnetization)^2 $,
where the final magnetization $\magnetization = \frac{1}{2} (\bar
n^\final_{\electron +} - \bar n^\final_{\electron -}$) is a universal
function of $g_\electron B/\Tk$. 
(At very large fields, however, a bulk Zeeman field, neglected above,
will spoil universality, see \SOMDIS~8.)  Thus, 
the exponents $\eta_\sigma (B)$ are \emph{universal functions of
  $g_\electron B/\Tk$}, with simple limits for small and large fields
(see Fig.~5b):
\begin{equation}
\eta_{\lowerlevel / \upperlevel}  \to 
\left\{
\begin{array}{ll}
\frac{1}{2} & (|B| \ll \Tk) \; ,  \rule[-4mm]{0mm}{0mm}
\\
\pm 1 & (|B| \gg \Tk) \;  .
\end{array}
\right. 
\label{eq:eta-m-limits}
\end{equation}  
The notation $\sigma = $ ``lower'' or ''upper'' distinguishes whether
the spin of the photo-excited electron (selectable by choice of
circular polarization of the incident light) matches the spin of the
lower or upper of the Zeeman-split e-levels, respectively.  The sign
difference between $\eta_{\lowerlevel}$ and $\eta_{\upperlevel}$ for
$|B| \gg \Tk $ arises since the change in local charge becomes fully
asymmetric, $\Delta n_{\electron , \lowerlevel } \to 1$ while $\Delta
n_{\electron , \upperlevel } \to 0$; as a result, for $\sigma =
$~lower Anderson orthogonality is completely absent ($\Delta
n'_{\electron \sigma'} = 0$), whereas for $\sigma = $~upper it is
maximal ($\Delta n'_{\electron \sigma'} = 1$).
It follows, remarkably, that a magnetic field tunes the strength of
Anderson orthogonality, implying a dramatic $\sigma$-dependence of the
evolution of the lineshape $A_\sigma (\nu ) \propto
\nu^{-\eta_\sigma}$ with increasing $|B|$ (Fig.~5a): For
$A_\lowerlevel (\nu)$, the near-threshold singularity becomes
stronger, tending towards $\nu^{-1}$. In contrast, for $A_\upperlevel
(\nu)$ the singularity becomes weaker, and once $\eta_\upperlevel$
turns negative, changes to an increasingly strong power-law decay,
tending toward $\nu^{+1}$; this is accompanied by the emergence of an
absorption peak near $\nu = |g_\electron B|$, associated with a
transition into the upper Zeeman-split level, broadened by Korringa
relaxation of its spin (see \SOMDIS~7). The fact that for $|B| >
\Tk $ a change of spin orientation of the photo-excited electron will
turn a near-threshold divergence in the lineshape to a suppression
that constitutes the low-frequency side of a broadened peak, is one of
the most striking predictions of our analysis.

\paragraph*{\bf Magnetic field dependence-dependence of the absorption threshold.}
The \emph{shift} of the absorption threshold frequency 
$\omega_\threshold = E^\final_\ground - E^\initial_\ground$
with magnetic field can be written as $\omega_\threshold (B) -
\omega_\threshold (0) = \frac{3}{2} \bar \sigma g_\hole B + \delta
\omega_\threshold^\electron (B)$.  The first term reflects the Zeeman
energy of the photo-excited hole (which has pseudo-spin 3/2), the 
second the $B$-dependence of the ground-state energy of the electron system.
The general $T=0$ relation
$g_\electron m^\aaaa_\electron =  \partial E_\ground^\aaaa / \partial B$
implies that the differential threshold shift offers a direct way of
experimentally measuring the local moment difference between the final
and initial ground states: $\partial (\delta
\omega_\threshold^\electron) / \partial B = g_\electron
[m_\electron^\final (B) - m_\electron^\initial (B) ]$.  Also, for
$\bar n^\initial_\electron \simeq 0$ and $\bar n^\final_\electron
\simeq 1$, where Eq.~(9) applies, this quantity can be related to the
infrared singularity exponents, allowing for a consistency check.
Moreover, the asymptotic behavior of $m^\final_\electron$ for small
fields ($\magnetization = - g_\electron B \chi_0$, where $\chi_0 = 1 /
4 \Tk$ is the linear static susceptibility) and large field
($|\magnetization| = \frac{1}{2}$) implies:
\begin{equation}
  \label{eq:thresholdshift}
  \delta \omega^\electron_\threshold = 
  \left\{ 
\begin{array}{ll}
- (g_\electron B)^2 / 8\Tk  \quad & ( |B| \ll \Tk) \; ,  \\
 - g_\electron B/2  &  (\Tk \ll B \ll |\varepsilon^\final_\electron|) \; . 
\end{array}
\right. 
\end{equation}
We note that a formula that interpolates through both of these regimes
is given by $\delta \omega^\electron_\threshold = \Tk -\frac{1}{2}
\sqrt{(g_\electron B)^2 + (2 \Tk^{2})} $ which, up to numerical
prefactors, has a similar functional form as the expression for the
ground state energy of the s-d model derived in Ref.~\cite{Ishii1968}.
The quadratic $B$-dependence of $ \delta \omega^\electron_\threshold $
for small fields offers a straightforward way to determine the Kondo
temperature experimentally. The accessibility of the $B$-dependence of
the ground state energy and e-level magnetization via the absorption
threshold is a remarkable advantage of the proposed optical probe of
Kondo physics in this paper -- these quantities are not accessible via
transport measurements.


We have demonstrated that optical absorption in a single quantum dot can be used to implement a quantum quench - a sudden change in the Hamiltonian governing the dynamics of the many-body system. Given that the relevant quantum dot parameters are tunable via external electric and magnetic fields, this system constitutes one of the rare, if not unique, experimentally accessible solid-state systems where a tunable quantum quench could be realized. Our work sets the stage for exploring numerous further interesting problems, such as (i) the effect of an exchange interaction between e- and h-levels; (ii) the effect of nuclear spins on the electron spin dynamics; (iii) using a pump-probe protocol to study the non-equilibrium time evolution even more directly; (iv) studying the
coherence of optical Raman transitions, where the virtually excited intermediate state is a Kondo correlated state of the QD electron and an adjacent fermionic reservoir, and (v) attempting to exploit strong correlations to realize a ``Fermionic quantum bus" between two distant QD spins. The presented results have an important bearing on the quantum control of a spin degree of freedom in nano-structures and on quantum optical techniques used to study reservoirs composed of fermionic constituents.

\section*{Methods}

\subsection{Numerical Renormalization Group for calculation of optical absorption lineshape}
The optical absorption lineshape given by Fermi's golden rule \Eq{eqabslsexact}, can be calculated at finite temperatures using full density matrix (FDM) numerical numerical renormalization group (NRG)\cite{Weichselbaum2007}. Because \Eq{eqabslsexact} contains matrix elements between initial and final eigenstates of different Hamiltonians, $H^\initial$ and $H^\final$, two separate NRG runs (NRG run \#1 and \#2) are required to calculate the initial and final eigenstates ($\{ |m\rangle_\initial \}$ and $\{ |n\rangle_\final \}$) as well as eigenenergies ($\{ E_m^\initial \}$ and $\{ E_m^\final \}$). The double sum in  \Eq{eqabslsexact}, over all initial and final eigenstates, is performed via a ``backwards'' run from the end to the beginning of the Wilson chain\cite{Anders2005}: for each shell $k$, the contribution towards the initial density matrix $\rho^\initial $ from that shell (obtained using data from NRG run \#1), and the matrix elements $\left| {}_\final \langle n | e^\dagger_\sigma | m \rangle_\initial \right|^2$ between shell-$k$ eigenstates from NRG runs \#2 and NRG \#1 are calculated, and binned  according to the corresponding frequency difference $E_n^\final - E^\initial_m$. See \SOMDIS~2 for further details.

\subsection{Non-equilibrium dynamics via NRG}
The expectation value of an observable $\hat B$ after absorption is given by $\widetilde B (t) = \mathrm{Tr} \left( \hat \rho^\final_\projected (t)  \hat B \right)$ where the time evolution is governed by the final Hamiltonian, $\hat \rho^\final_\projected (t) \equiv e^{-i H^\final t} \hat \rho^\final_\projected e^{iH^\final t}$. For reasons discussed in  \SOMDIS~3 we find it convenient to take $\hat \rho^\final_\projected  = \hat \rho_\projected / [ 1 - \bar n^\initial_{\electron \sigma} ]$. The Fourier transform of $\widetilde B (t)$, $ \widetilde {\cal B} (\omega) = \int dt~e^{i\omega t} \widetilde B (t) $, can be expressed in Lehmann representation:
\begin{eqnarray}
\widetilde {\cal B} (\omega)  =
\sum_{n,n^{\prime
}} {}_\final \! \langle n' | \hat \rho^\final_\projected | n \rangle_\final 
 {}_\final \! \langle n | \hat{B} | n' \rangle_\final 
\cdot 2\pi \delta \left( \omega
- E_{n'}^\final + E_{n}^\final \right) \text{.}
\end{eqnarray}%
This expression can be calculated using FDM-NRG. See \SOMDIS~3 for details.

\subsection{Fixed-point perturbation theory}

To calculate $A_\sigma (\nu)$ at $T=0$ for $\nu \gtrsim T_{K}$, set $H^\final \to H_\regime^\ast + H^\prime_\regime $ in \Eq{eq:T=0Gee} and expand in powers of $H^\prime_\regime$, for $\regime = \FO$ or $\LM$. At zeroth order this yields a peak $A_\sigma (\nu) \sim \delta (\nu)$, which is irrelevant for $\nu \gtrsim T_{K}$.  To lowest non vanishing
order in $H^\prime_\regime$
\begin{equation}
A^\regime_\sigma (\nu) = 
- \frac{2}{\nu^2} {\rm Im} \;
\Bigl[ {}_\initial \langle \ground |
e^\pdag_\sigma \, H^\prime_\regime
 \frac{1}{\nu_+ - \barH^\ast_\regime }
H^\prime_\regime \,
e^\dagger_\sigma
| \ground \rangle_\initial \Bigr] , \qphan
\label{eq:A-FO-LM-startingpoint}
\end{equation}
which we shall evaluate for $T=B=0$, $|\efinal| = \frac{1}{2} \Uee$
and $|\ground \rangle_\initial \simeq \prod_{\varepsilon_{k\sigma} <
  \varepsilon_\Fermi} c_{k\sigma}^\dagger |{\rm Vac} \rangle$.  We find
\begin{equation}
A^\regime_\sigma (\nu) = - \frac{2}{\nu^2}\im{ \; T_{\sigma\sigma}^{\regime}(\nu)} \; , 
\end{equation}
where $T_{\sigma\sigma}^{\FO}(\nu) = 2 \, (
\Gamma/\pi \rho ) \, {\cal F}_{\nu} \left\{ -i \theta (t) \, {}_\initial \langle
  \ground | c_\sigma (t)c^\dag_\sigma (0)| \ground \rangle_\initial
\right\}$ in the FO-regime (${\cal F}_{\nu}$ is the Fourier-transform operator ${\cal F}_{\nu} \left\{ A \right\} \equiv \int {\rm d} t \, e^{it(\nu_+ + \omega_\threshold)} A(t)$). In this regime, the absorption process can be understood
as a two-step process consisting of a virtual excitation of the QD
resonance, followed by a tunneling event to a final free-electron
state above the Fermi-level. Therefore, the absorption rate is
proportional to the tunneling density of states (into the FR) at
detuning $\nu$. On the other hand, in the LM regime,
$T_{\sigma\sigma}^{\LM}(\nu) = 3/8 \, (J/\rho)^{2} \, {\cal F}_{\nu}
\left\{ -i \theta (t)\, {}_\initial \langle \ground | s_{c}^{\sigma} (t)
  s_{c}^{-\sigma} (0) | \ground \rangle_\initial \right\}$ where we
used $\langle s_{c}^{z}(t) s_{c}^{z}(0) \rangle = \langle s_{c}^{+}(t)
s_{c}^{-}(0) \rangle / 2$. This result reiterates that the
intermediate detuning probes spin fluctuations, as observed in the
dynamics (Fig.~2). Evaluating these correlation functions yields \Eq{eq:A-FO-results} and \Eq{eq:A-LM-results}. In Eq.~(\ref{eq:A-LM-results}), we have inserted the scale-dependent exchange interaction $J(\nu) = \ln^{-1}(\nu/\Tk)$ which results from
the logarithmic enhancement of the exchange interaction within perturbative RG. The finite-temperature form ($T \gg \Tk$) of the lineshape is discussed in \SOMDIS~5 (see also Fig.~\ref{fig:finiteTsom}).

\subsection{Absorption in the strong-coupling regime and Fermi-edge physics}

\Eq{eq:T=0Gee} can be expressed as the Fourier transform,  ${\cal G}_{\electron \electron}^\sigma (\nu) ={\cal F}_{\nu} \left\{ G_{\electron \electron}^\sigma (t) \right\}$, of a correlator 
\begin{eqnarray}
  G_{\electron \electron}^\sigma (t) & = & - i \theta (t) \,
\langle
e_\sigma (t) e_\sigma^\dagger
\rangle_\initial \; , 
\end{eqnarray}
involving operators defined to have an anomalous time dependence,
$\hat O(t) = e^{i H^\initial t} \hat O e^{-i H^\final t}$.  This
anomalous time dependence, involving both $H^\initial$ and $H^\final$,
reflects the fact that the creation of a hole during optical
absorption abruptly lowers the \electron-level. We next relate ${\cal G}_{\electron \electron}^\sigma (\nu)$ to a similarly-defined correlator of FR electrons, 
\begin{equation}
{\cal G}_{k k'}^\sigma (\nu) =  {\cal F}_{\nu} \left\{ - i \theta (t) \, \langle c_{k \sigma} (t) c_{k' \sigma}^\dag
\rangle_\initial \right\}
\end{equation}
using equations of motion in the asymptotic limit $\nu \rightarrow 0$:
\begin{eqnarray}
{\cal G}_{k k'}^\sigma (\nu) \sim 
\frac{v^2 {\cal G}^\sigma_{\electron \electron} (\nu)}{
(\nu_+ + \Delta - 
\varepsilon_{k \sigma})
(\nu_+ + \Delta - 
\varepsilon_{k' \sigma}) } \; , 
\end{eqnarray}
where $\Delta = \omega_\threshold - \varepsilon_{\hole \bar \sigma} $. This asymptotic relation is established in \SOMDIS~6 and  implies 
\begin{eqnarray}
\label{eq:AGccM}
A^\SC_\sigma (\nu) \sim  \frac{2}{\pi \rho \Gamma}
 {\rm Im} {\cal G}_{\FermiR \FermiR}^\sigma (\nu)  \;  .
\end{eqnarray}
where ${\cal G}^\sigma_{\FermiR \FermiR} (\nu) = \sum_{kk'} {\cal G}_{k k'}^\sigma (\nu)$. To calculate $G^\sigma_{\FermiR \FermiR}(t)$ for $t \gg 1/\Tk$, we may
now replace $H^\initial \to H_{\FermiR}$ and $H^\final \to
H^\ast_\SC + H^\prime_\SC$ in Eq.~(\ref{eq:AGccM}):
\begin{equation}
  G_{\FermiR \FermiR}^{\sigma}(t)
\sim 
{}_\initial \langle \ground | \ex{iH_\FermiR t}
c_{\sigma} \ex{-iH_{\SC}^\ast t}
c_{\sigma}^\dagger
| \ground \rangle_\initial
\; .
\end{equation}
This response function is similar to that calculated in the X-ray edge problem\cite{Mahan1967,Nozieres1969,Ambrumenil2005} and its calculation is standard  (e.g. \cite{Ambrumenil2005}) and yields (we show only the leading power law)
\begin{eqnarray}
G^\sigma_{\FermiR \FermiR} (t) & \sim &
 t^{-[(\delta_\sigma - \pi)^2 + \delta^2_{\bar \sigma}]/\pi^2}
  \; . \qphan
\end{eqnarray}
where $\delta_\sigma = \delta_\sigma (0)$ denotes the phase shifts at the Fermi energy. The phase shifts at the Fermi energy,
in turn, are given by $\delta_\sigma  = \pi \Delta n_{\electron \sigma} $, according to the Friedel sum rule\cite{Friedel1956}, valid for $T=0$ and for arbitrary values of $B$, $\bar n^\final_\electron $ and $\bar n^\initial_\electron
$. Collecting results, we find \Eq{eq:ASC} and \Eq{eq:exponentsmalldetuning}.  See \SOMDIS~6 for details.

\begin{acknowledgments}
AI and HET acknowledge support from the Swiss NSF under Grant
  No. 200021-121757. HET acknowledges support from the Swiss NSF under
  Grant No. PP00P2-123519/1. BB acknowledges support from the Swiss NSF and NCCR Nanoscience (Basel). JvD acknowledges support from the DFG (SFB631, SFB-TR12, De730/3-2, De730/4-1), the Cluster of Excellence
  \emph{Nanosystems Initiative Munich} and in part the National
  Science Foundation under Grant No.  NSF PHY05-51164.   AI acknowledges
  support from an ERC Advanced Investigator Grant, and LG from NSF Grant No. 
  DMR-0754613.
\end{acknowledgments}

\setcounter{equation}{0}
\setcounter{figure}{0}
\renewcommand{\theequation}{S\arabic{equation}}
\renewcommand{\thefigure}{S\arabic{figure}}

\begin{appendix}
\section*{Appendix 1: The model and its experimental realization} 

\begin{figure}
\centering
\includegraphics[width=\linewidth]{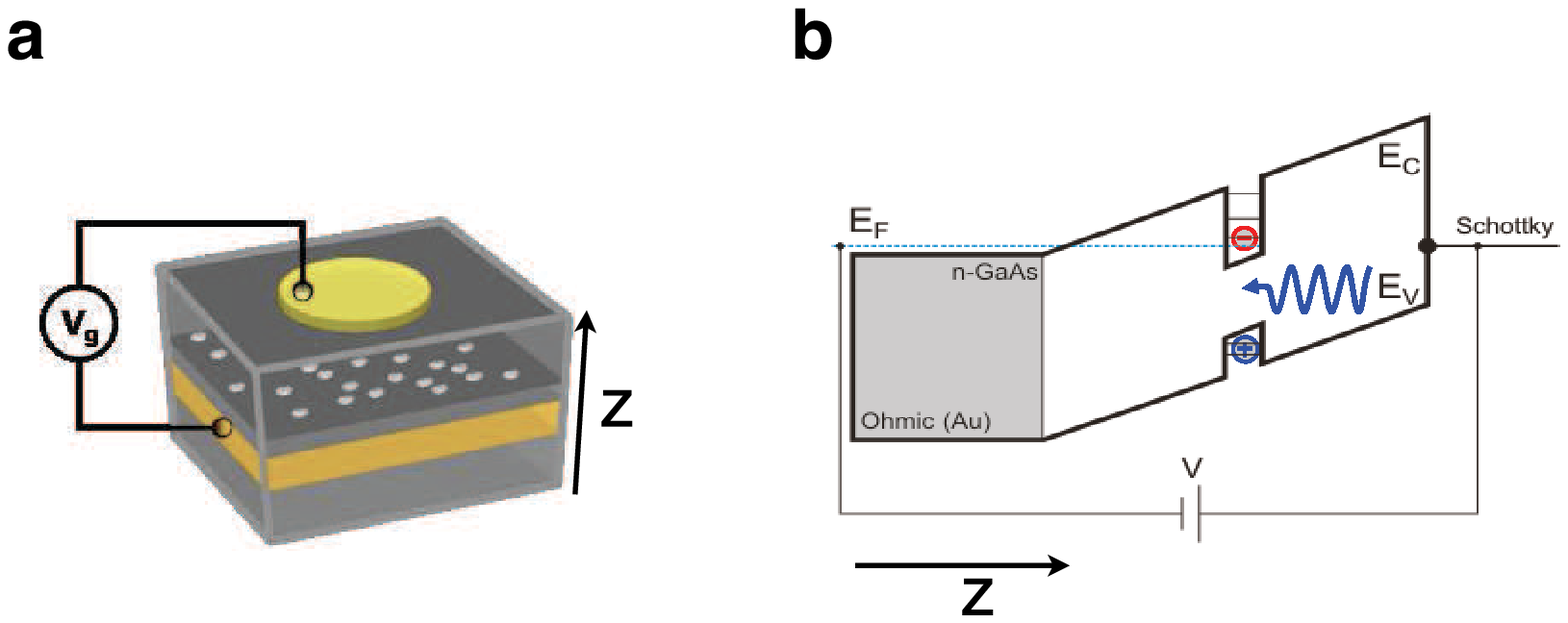}
\caption{{\bf The Schottky diode structure for the experimental observation of the optical signatures of the Kondo effect}.}
\label{fig:S1}
\end{figure}

We focus here on differential transmission spectroscopy of single
semiconductor QDs in gated heterostructures with an adjacent
tunnel-coupled electron reservoir. A possible realization is a
Schottky diode structure{\Hoegele} where a voltage applied between a
top Schottky contact and a two-dimensional electron gas (2DEG) in a
modulation-doped GaAs layer underneath a layer of self-assembled InAs
QDs, is used to adjust the relative energy of the QD electron with
respect to the Fermi-energy of the 2DEG (Figure~\ref{fig:S1}). To realize the
situation in Fig.~1 we consider here a range of gate voltages for
which the QD is uncharged before the optical excitation (See Fig.~4,
upper inset). In contrast to most of the earlier quantum optics
experiments where the separation between the electron gas and the QD
layer was kept large to ensure weak spin-flip co-tunneling, we focus
here on structures with a small barrier where tunnel coupling of a QD
electron is only a factor of 10 weaker than its single-electron
charging energy. We estimate that with a 15 nm neutral GaAs barrier,
the strength of the exchange interactions would be strong enough to
yield Kondo temperatures exceeding 100 mK.

Spin-orbit interaction in InGaAs QDs leads to a splitting of about 200
meV between the $J=1/2$ and $J=3/2$ valence band states. In the limit
of asymmetric quantum confinement typical for our QDs, the four
$J=3/2$ bands further split into two Kramers doublets. The states with
$J_z = \pm 3/2$ along the growth (= strong confinement) direction have
the lowest zero-point energy due to their heavy mass for motion along
the growth direction and hence are the ones relevant for lowest energy
optical excitations. The confinement+strain induced splitting between
the $J_z = \pm 3/2$ (heavy-hole) and $J_z = \pm 1/2$ (light-hole)
states is estimated to be around 20 meV. In the main text, the spin
label of the valence band electrons (or holes, $h_{\sigma}$) therefore
refer to pseudo-spin $J_{z} = \pm 3/2$, and the
hole Zeeman energy has the form $\sigma \frac{3}{2} g_\hole B$. 

This situation leads to near perfect correlation between the circular
polarization of the excitation light and the spin of the optically
excited electron in InGaAs QDs (as well as any QD structure with
possibly the exception of chemically synthesized nanocrystals). The
hole g-factor in these QDs is highly anisotropic: along the growth
direction, the value is around $1.2$. In plane g-factor ranges from
$0-0.5$.

For the calculations in the main body of the paper, we assume the
electron-hole recombination rate (typically of order $\Gamma_{\rm eh} \approx
1\mu$eV) to be negligible compared to all other scales, take the hole
level to lie within the gap of the FR and neglect its coupling to the
latter. The newly-created electron-hole pair will experience a mutual
Coulomb attraction $\Ueh (>0)$ and a much weaker exchange interaction
$\Jeh$ (typically of order $\Jeh \approx 200$ $\mu$eV){\Tureci}.

While the interplay of the intra-dot electron-hole exchange
interaction and the Kondo correlations is interesting on its own
right, the effects of electron-hole exchange can experimentally be
avoided in two ways. One possible alternative is to start out with a
single-hole charged QD; the experiments on single QDs embedded in
n-type Schottky structures have already shown that the lifetime of
such an optically charged state could well exceed $100$
$\mu$sec. Resonant optical excitation of such a single-hole charged QD
leads to the formation of the so-called $X^{1+}$ trion which consists
of two valence-band heavy-holes forming a singlet and a single e-level
electron with vanishing electron-hole exchange interaction. Another
possible alternative is using indirect excitons in coupled QDs where
the electron and the hole wave-functions have vanishing spatial
overlap.

Here, we assume $\Jeh \ll T_{K}$, or an experimental realization where
it can neglected, so that the hole has no dynamics after
creation. With the hole ``frozen'', the electron dynamics can be
described by a pure Anderson model (AM), Eq.~(1), for the e-level
(involving only $e_\sigma$, not $h_{\bar \sigma}$), whose initial and
final energies before and after absorption,
$\varepsilon^{\aaaa}_{\electron \sigma}$ ($\aaaa = \initial, \final$),
are related by $\varepsilon^\aaaa_{\electron \sigma} =
\varepsilon_{\electron \sigma} - \delta_{\aaaa \final} \Ueh$
(Fig.~1b).  The downward pull of the Coulomb attraction $\Ueh$
stabilizes the excited electron against decay into the FR, provided
that $\efinal $ lies below the FR's Fermi energy $\varepsilon_\Fermi =
0 $.

\section*{Appendix 2: Numerical Renormalization Group and Fermi's Golden Rule}

The numerical renormalization group{\Krishna} is an iterative
method for numerically diagonalizing quantum impurity models such as
Anderson impurity Hamiltonians $H^\aaaa$ ($\aaaa = \initial, \final$)
specified around Eq.~(1).
The spectrum of states of the Fermi reservoir is coarse-grained using
a logarithmic discretization scheme governed by a parameter $\Lambda >
1$ (typically $\Lambda=2$), followed by an exact mapping of the
discretized model onto a semi-infinite chain, the so-called Wilson
chain, whose hopping amplitudes decay exponentially along the chain,
as $t_{k}\sim \Lambda^{-k/2}$. This produces a separation of energy
scales and makes it possible to diagonalize the Hamiltonian
iteratively: knowing the eigenstates of a chain of length $k-1$, one
adds site $k$ and calculates the ``shell'' of eigenenergies of the
Hamiltonian for the chain of lenght $k$. The high-lying eigenstates of
that shell are ``discarded'', while the low-lying states are ``kept''
and used for the next iteration.  The spectrum of eigenenergies so
obtained typically flows past one or more non-stable fixed-points and
finally converges towards a stable fixed point, whereupon the
iterative procedure can be stopped. In practice one thus deals with a
finite Wilson chain, whose length is set by the smallest energy scale
in the system (e.g.\ the Kondo temperature, temperature, or magnetic
field). By combining NRG data from all iterations, it is possible to
construct a \emph{complete} set{\Anders} of approximate many-body
eigenstates of the full Hamiltonian. These can be used to evaluate
equilibrium spectral functions via their Lehmann-representations; at
finite temperatures, this can be done using the full density matrix (FDM)-NRG{\Weichselbaum}.


Since Eq.~({\eqFGR}) expresses the Fermi golden rule absorption rate via a
Lehmann representation, it, too, can be evaluated using NRG{\Helmes}. However, it contains matrix elements between initial and
final states that are eigenstates of different Hamiltonians,
$H^\initial$ and $H^\final$.  Hence, two separate NRG runs are
required to calculate these (similar in spirit to what is done for
time-dependent NRG{\Anders}). The strategy is then as follows:
 
\begin{itemize}
\item NRG run \#1 generates a complete set of approximate eigenstates
$|m\rangle_\initial$ and eigenenergies $E_m^\initial$ 
for the intial Hamiltonian $H^\initial$ (without exciton).
\item NRG run \#2 generates a complete set of approximate eigenstates
$|n\rangle_\final$ and eigenenergies $E_m^\final$ 
for the final Hamiltonian $H^\final$ (with exciton).
\item The double sum in Eq.~({\eqFGR}), over all initial and final
  eigenstates, is performed via a ``backwards'' run, with site
 index $k$ running from the end to
  the beginning of the Wilson chain{\Anders}: for each shell $k$,
   the contribution towards
  the initial density matrix $\rho^\initial $ from that shell
  (obtained using data from NRG run \#1), and the matrix elements
  $\left| {}_\final \langle n | e^\dagger_\sigma | m \rangle_\initial
  \right|^2$ between shell-$k$ eigenstates from NRG runs \#2
  and NRG \#1 are calculated, and binned (see below) according to the
  corresponding frequency difference $E_n^\final - E^\initial_m$.
\item The $T=0$ threshold frequency for the onset of absorption is
  given by the difference of groundstate energies of NRG runs \#2 and
  \#1, $\omega_\threshold \equiv E_\ground^\final
  -E_\ground^\initial$.  The absorption spectrum is expected to have
  divergences at the threshold $\omega_\threshold$, hence all
  frequency data are shifted by the overall threshold energy
  $\omega_\threshold $ prior to binning.  (For finite temperature, the
  sharp onset is broadened and divergencies are cut off.)
\item The discrete eigenenergies of shell $k$ are spread over an
  energy range comparable to the characteristic energy
  $\Lambda^{-k/2}$ scale of that iteration, which decreases
  exponentially with $k$.  Thus, the bins used for collecting the
  discrete data are likewise chosen to have widths decreasing
  exponentially with decreasing energy. The discrete, binned data are
  subsequently broadened using a log-Gaussian broadening scheme,
  characterized by a broadening parameter $\alpha${\Weichselbaum},
  typically taken as $\alpha = 0.6$. For finite temperature, the
  maximum of the absorption peak occurs at a frequency $\nu_{\rm
    max}$, that is slightly larger than the $T=0$ threshold frequency
  at $\nu = 0$ (see Fig.~3a).  Thus, for finite temperature, the
  binning procedures (both the binning frequencies and the frequency
  ranges in which we change from log-Gaussian to Gaussian) for
  frequencies above and below $\nu_{\rm max}$ were set up to be
  symmetric with respect to $\nu_{\rm max}$.  This ensures that the
  $A_\sigma (\nu)$ curves for $\nu$ smaller and larger than $\nu_{\rm
    max}$ match smoothly.

\end{itemize}

\section*{Appendix 3: Time evolution via NRG}

The nonequilibrium time evolution of the system in response to the
sudden creation of an $e^\dagger_\sigma h^\dagger_{\bar \sigma}$
exciton at time $t=0$ can be calculated using time-dependent NRG{\Anders}.  The sudden addition of a $\sigma$-electron in the e-level
projects the system's initial density matrix $\hat \rho^\initial$ onto
a projected density matrix $\hat \rho_\projected$,
\begin{eqnarray}
\hat{\rho}^\initial \equiv
\sum_{m} |m\rangle_\initial (\rho_{m}^\initial) \, {}_\initial\! \langle m | 
\rightarrow
\hat{\rho}_\projected \equiv 
e^\dagger_\sigma \hat \rho^\initial  e_\sigma \; . 
\end{eqnarray}
Its norm is smaller than 1,
\begin{eqnarray}
\label{eq:normrhofinal}
{\rm Tr} \hat \rho_\projected = 
{\rm Tr} \left( \hat \rho^\initial e_\sigma e^\dagger_\sigma \right) 
= 1 - \bar n^\initial_{\electron \sigma} \; ,
\end{eqnarray}
 reflecting  the fact that states in which the e-level
had already contained a $\sigma$-electron prior to absorption are
projected to 0. ($\bar n^\initial_{\electron \sigma} = \langle \hat n_{\electron
  \sigma} \rangle_\initial$ denotes an expectation value w.r.t.\ to
$H^\initial$.) After this projection, the e-level occupancies for
spin $\sigma'$ are reset (or ``reinitialized'') to be
\begin{eqnarray}
  \label{eq:initialn_esigma}
{\rm Tr} 
( \hat \rho_\projected \hat n_{\electron \sigma'})
 = \left\{ \begin{array}{ll} 
  \bar n^\initial_{\electron,00} +   \bar n^\initial_{\electron,0 \bar \sigma}  = 
  1 - \bar n^\initial_{\electron \sigma} & \, 
 \textrm{for} \; \sigma' = \sigma \;  
\\
  \bar n^\initial_{\electron,0\bar \sigma} & \, 
 \textrm{for} \; \sigma' = \bar \sigma  \; 
\end{array} \right. 
\end{eqnarray}
where $  \bar n^\initial_{\electron,00}$ and $\bar n^\initial_{\electron,0\bar \sigma}$
are the initial probabilities (with respect to $H^\initial$) 
for the e-level to have been completely empty $(00)$,
or to have contained no $\sigma$-electron but an $\bar \sigma$-electron
$(0 \bar \sigma)$,  respectively.  

For the sake of studying the time evolution after absorption, we
find it convenient to take $\hat \rho^\final_\projected  =
\hat \rho_\projected / [ 1 - \bar n^\initial_{\electron \sigma} ]$ as
starting density matrix, normalized such that ${\rm Tr} \hat
\rho^\final_\projected  = 1$ and ${\rm Tr} \hat
\rho^\final_\projected  \hat n_{\electron \sigma} = 1$.  Its
subsequent time evolution is governed by the final Hamiltonian, $\hat
\rho^\final_\projected (t) \equiv e^{-i H^\final t} \hat
\rho^\final_\projected e^{iH^\final t}$. The corresponding expectation
value of an observable $\hat B$ is thus given by
\begin{equation}
\widetilde B (t) = 
\mathrm{Tr} \left(
\hat \rho^\final_\projected (t)  \hat B \right) \; . \label{fgr_tdep}%
\end{equation}
Fourier-transformed to frequency space using $\widetilde {\cal B} (\omega) =
\int dt~e^{i\omega t} \widetilde B (t) $, one obtains
\begin{eqnarray}
\widetilde {\cal B} (\omega)  =
\sum_{n,n^{\prime
}} {}_\final \! \langle n' | \hat \rho^\final_\projected | n \rangle_\final 
 {}_\final \! \langle n | \hat{B} | n' \rangle_\final 
\cdot 2\pi \delta \left( \omega
- E_{n'}^\final + E_{n}^\final \right) \text{.}  \label{fgr_tdep_real}
\end{eqnarray}%
The latter expression, being written in Lehmann representation, is
again well-suited for FDM-NRG{\Weichselbaum}. Following
{\Anders}, it is again possible to formulate the procedure such
that matrix elements are always calculated within the same Wilson
shell.

The discrete data contributing to $\widetilde {\cal B} (\omega)$
is binned and smoothened using a log-Gaussian broadening function,
as described in the previous section, but using 
a smaller broadening parameter, $\alpha = 0.3$.
$\widetilde B(t)$ is obtained by
  Fourier-transforming the broadened version of $\tilde B(\omega)$ to
  the time domain at the very end of the calculation.

  The values obtained for $\widetilde n_{\electron} (\infty)$,
  $\widetilde n_{\electron\sigma} (\infty)$ and $\widetilde
  m_\electron (\infty)$ deviate from their expected equilibrium
  values (1, $\frac{1}{2}$ and 0, respectively, for the case
  $H^\final$=SEAM depicted in Fig.~2) by about 3~\%. This is a known
  artifact of time-dependent NRG{\Anders}, presumably due to the fact that the
  NRG discretization scheme inevitably is rather coarse at large
  energies.

\section*{Appendix 4: Perturbation 
theory for the excitonic Anderson Model}

The optical absorption lineshape for a $\sigma=+$ excitation and the
excitonic Anderson Model can be calculated via a perturbation
expansion up to second order in the tunnel coupling
$v=\sqrt{\Gamma/\pi \rho}$ (we keep here the momentum dependence of
$v_{k}$) for the initial and final states entering Eq.~(2).
(The wavefunctions used below are similar in structure
to those used in Ref.{\Gunnarsson} for variational
calculations of various emission and absorption spectra of Ce
compounds.)

We define the empty e-level, h-level and unperturbed FR as $\bra{0} =
\bra{0}_\electron \otimes \bra{0}_\hole \otimes \bra{0}_F$. The bare
excitonic resonance frequency is $\omega_0 = \omega_\threshold =
\varepsilon_\electron - \Ueh +  \varepsilon_\hole $ which
also marks the approximate onset of the edge (threshold) behavior. We
write the initial state $| \ground \rangle_\initial$ expanded to
second order in $v_k$ as
\begin{align}
  | \ground \rangle_\initial = [ \,  1 +  &
    \sum_{\substack{k<0\\ \sigma}} \frac{v_k}{\varepsilon_k-\varepsilon_\electron } \esigD\csigk 
   &  \nonumber \\
   &  + \sum_{\substack{k<0\\ k'>0\\ \sigma}} \frac{v_k v_{k'}^*}{(\varepsilon_k-\varepsilon_{k'})(\varepsilon_\electron -\varepsilon_k)} \csigqD\csigk \, ] \bra{0}
\end{align}
The set of final states are defined by particle number conservation. A
few possible zeroth order final states accessible from the initial
state (to low orders in $v_{k}$) are: $\csigkD\hdD\bra{0}$ ($X^{+}$
exciton), $\edD\cdk\euD\hdD\bra{0}$ ($X^-$ exciton) and
$\csigkD\hat{c}_{\vec{k'},\sigma'} \EmuD\hdD\bra{0}$ ($X^{0}$ + e-h
excitation). Note that the last class of states includes for $k=k'$
the bare neutral exciton final state. The relevant final states are
then
\begin{align} 
	\bra{X^{+}_{k\sigma}} &= \left[ 1 + \frac{v_k}{\varepsilon_k-\varepsilon_\electron +\Ueh } \esigD\csigk \right. \nonumber \\ 
	& \left. + \sum_{k'} \frac{v_{k'}}{\varepsilon_{k'}-\varepsilon_\electron +\Ueh } \esignD\csignq \right] \csigkD\hdD \bra{0}  \\
	\bra{X^-_k} &= \left[ 1 + \frac{v_k^*}{\varepsilon_\electron +\Uee -\Ueh -\varepsilon_k} \cdkD\edsi \right. \nonumber \\
	& \left. + \sum_{k'} \frac{v_{k'}^*}{\varepsilon_\electron +\Uee -\Ueh -\varepsilon_{k'}}\cukD\eusi \right] \edD\cdk\euD\hdD\bra{0} \\
	\bra{X^{0}_{k\sigma k' \sigma' \mu}} = &\left[ 1 + \delta_{\sigma',\mu}\frac{v_{k'}^*}{\varepsilon_\electron -\Ueh -\varepsilon_{k'}}\cnuqD\Enu   \right. \nonumber \\
	& \left. + \delta_{\sigma,-\mu}\frac{v_k}{\varepsilon_k-\varepsilon_\electron -\Uee +\Ueh }\EnunD\cnunk \right. \notag\\
	&\left. + \delta_{\sigma',\mu}\frac{v_k v_{k'}^*}{(\varepsilon_k-\varepsilon_{k'})(\varepsilon_\electron -\Ueh -\varepsilon_{k'})} \esigD\csigk\cnuqD\Enu \right. \notag\\
	& + \delta_{\sigma,-\mu} \sum_{q\mu} \frac{v_k v_q^*}{(\varepsilon_k-\varepsilon_q)(\varepsilon_k-\varepsilon_\electron -\Uee +\Ueh )} \notag \\
	& \left. \times \, \cmuQD\Emu\EnunD\cnunk  \right] \csigkD\hat{c}_{k',\sigma'}\EnuD\hdD\bra{0} 
\end{align}
We have included the lowest order contributions in e-h excitations in
the FR. The final state energies to lowest order in $v_{k}$ are given
by
\begin{eqnarray} 
E_0^{(X^{+})} & = & \omega_\threshold + [
\varepsilon_k - (\varepsilon_\electron - \Ueh )] \; , 
\\
E_0^{(X^-)} & = &
\omega_\threshold + [(\varepsilon_\electron - \Ueh  + \Uee ) - \varepsilon_k] \; ,
\\
E_0^{(X^{0})} & = & \omega_\threshold + \varepsilon_k -
\varepsilon_{k'} \; .
\end{eqnarray}
 For SEAM, the analytical form of the lineshape is:
\begin{align}
  A_\uparrow(\nu) = & \frac{8\Gamma^{2}}{\pi} \left(
    \frac{1}{\nu}\frac{1}{U}\left[
      \frac{2}{\frac{U}{2}+\nu}+\frac{5}{\frac{U}{2}-\nu} \right] + \right. \nonumber \\
   &  \left. \frac{1}{\nu^3}
    \log\left[\frac{\left(\frac{U}{2}\right)^2-\nu^2}{\left(\frac{U}{2}\right)^2}\right]
    + \frac{1}{\nu^2}\frac{1}{U}
    \log\left[\frac{\frac{U}{2}+\nu}{\frac{U}{2}-\nu}\right]
  \right) \nonumber \\
  & + \frac{10\Gamma}{\nu^2} \cdot
  \theta(\nu - \frac{U}{2}) \label{eq:SEAMlineshape}
\end{align}

The singularity at $\nu=U/2$ can be related to a tunneling-assisted
$X^{-}$ resonance. The intermediate state there is an $X^{-}$ charging
state which is reached from the $X^{0}$ state by a FR electron
tunneling in, and thereby satisfying energy conservation.
We expect that a calculation including higher orders of perturbation
theory will regularize this singularity. 

The FO ``threshold'' term $ \theta(\nu - \frac{U}{2})$, which corresponds
to Eq.~(\ref{eq:A-FO-results}), is specific to the \emph{symmetric} Anderson model.  For the
asymmetric Anderson model, this threshold splits into two, at $\nu =
|\varepsilon^\final_{\electron \sigma}|$ and $\nu =
\varepsilon^\final_{\electron \sigma} + U$.

\section*{Appendix 5: Finite temperature fixed-point perturbation theory} 

\begin{figure}
\centering
\includegraphics[width=0.8\linewidth]{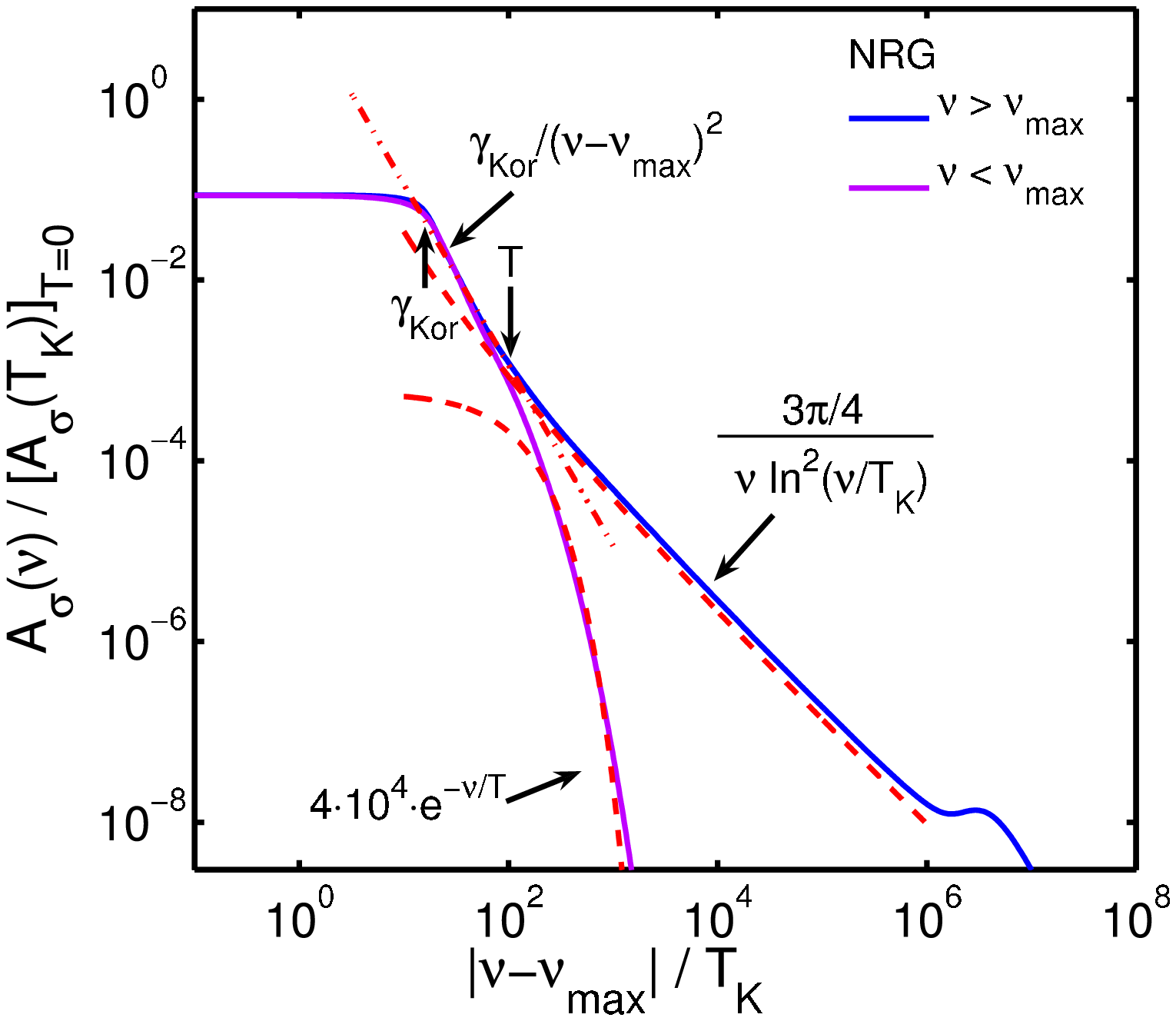}
\caption{{\bf The absorption lineshape at high temperatures}. For $T \gg\Tk$, the log-log plot shows the two portions of the absorption lineshape for red-detuning ($\nu< \nu_{\rm max}$) and blue-detuning ($\nu> \nu_{\rm max}$) with respect to the frequency $\nu_{\rm max}$ (at which $A_\sigma (\nu)$ reaches its maximum). Here, $\gamma_{\rm Kor} = 0.15 \, T$. NRG parameters: $\Uee=0.1D$, $\varepsilon^\initial_{\electron} = 0.75U$, $\varepsilon^\final_{\electron} = -0.5 \Uee$, $\Gamma=0.03 \Uee$, $\Tk = 5.9 \cdot 10^{-6} \, \Gamma$, $T= 100 \, \Tk$, $B=0$, $\Lambda = 1.8$, Kept states: $1024$, $\alpha = 0.5$.}
\label{fig:finiteTsom}
\end{figure}

To calculate $A_\sigma (\nu)$ for $T \gg \Tk$ set $H^\final \to H_\regime^\ast + H^\prime_\regime $ in Eq.~(3) and expand in powers of $H^\prime_\regime$, for $\regime = \FO$ or $\LM$. Note that the proper finite-T generalization of Eq.~(3) is ${\cal G}_{\electron \electron}^\sigma (\nu) = {\cal F}_{\nu} \left\{ - i \theta (t) \, \langle e_\sigma (t) e_\sigma^\dagger \rangle_\initial \right\}$. Let us focus only on the temperature dependence of the lineshape for $|\nu| < |\efinal|$ and $T \gg \Tk$:
\begin{equation}
  A_\sigma (\nu) =  \frac{3\pi/4}{1 - \ex{-\nu / T}} 
\frac{\gamma_{\rm Kor}(\nu,T)/\pi}{\nu^{2} + \gamma_{Kor}^{2}(\nu,T)}
\end{equation}
where $\gamma_{\rm Kor}(\nu,T)$ is the scale-dependent Korringa relaxation
rate given by
\begin{eqnarray}
\gamma_{\rm Kor}(\nu,T) = \left\{ \begin{array}{ll} 
 \pi T/\ln^{2} | T/\Tk | & \quad 
 \textrm{for} \;  |\nu| < T \; , 
\\
\pi \nu / \ln^{2} | \nu / \Tk | & \quad 
 \textrm{for} \; |\nu| > T  \;  
\end{array} \right. 
\end{eqnarray}
This expression is compared to the lineshape calculated by NRG in Fig.~\ref{fig:finiteTsom}. Note that for $\nu > T$, we recover Eq.~({\eqALM}).

\section*{Appendix 6: Evaluation of the absorption lineshape in the strong-coupling regime}

As we descend to detunings well below the Kondo scale, $\nu \ll T_K$,
the physics is governed by the strong coupling fixed point. It
describes a fully screened singlet, acting as source of strong
potential scattering for other FR electrons, causing the phase of each
mode $k\sigma$ to shift by $\delta_{\sigma} (\varepsilon_{k\sigma})$
relative to its value for $H^\initial$.  According to Nozi\`eres{\Nozieres} (see also \GlazmanPustilnik), the fixed
point Hamiltonian at $B=0$ is $H^\ast_\SC = \sum_{k\sigma}
\varepsilon_{k} \tilde c^\dagger_{k \sigma} \tilde c^\pdag_{k
  \sigma}$, where tildes denote operators representing phase-shifted
modes.  The leading relevant perturbation for $0 < B \lesssim \Tk$ has
the form\cite{Glazman2005,Nozieres1974} $H^\prime_\SC = \sum_\sigma \sigma
g_\electron B \, [ \, \sum_{k} {\textstyle \frac{1}{2}} \tilde
c^\dagger_{k \sigma} \tilde c^\pdag_{k \sigma} + \tilde \rho_\sigma
/(\pi \rho \Tk) \, ]$.

We shall not explicitly use this fixed point Hamiltonian, however,
since the strategy (described in Methods section) of
perturbing around the fixed point is of no use for calculating ${\cal
  G}_{\electron \electron}^\sigma (\nu)$ of Eq.~(3). The reason is that
${\cal G}_{\electron \electron}^\sigma (\nu)$ is formulated in terms
of $e_{\electron \sigma}$ and $e^\dagger_{\electron \sigma}$
operators, whose dynamics is determined by higher-energy excitations
of the FR \emph{not} described by $H^\ast_\SC + H^\prime_\SC$.  To
circumvent this problem, we use an equation of motion approach to
first derive an asymptotic relation between the impurity Green's
function and the Green's function of the itinerant electrons.

We start by noting that the correlator occuring in Eq.~(3) can
be expressed as the Fourier transform, 
${\cal G}_{\electron \electron}^\sigma (\nu) ={\cal F}_{\nu} \left\{ G_{\electron \electron}^\sigma (t) \right\}$, of a correlator 
\begin{eqnarray}
  \label{eq:GeetSOM}
  G_{\electron \electron}^\sigma (t) & = & - i \theta (t) \,
\langle
e_\sigma (t) e_\sigma^\dagger
\rangle_\initial \; , 
\end{eqnarray}
involving operators defined to have an anomalous time dependence,
$\hat O(t) = e^{i H^\initial t} \hat O e^{-i H^\final t}$.  This
anomalous time dependence, involving both $H^\initial$ and $H^\final$,
reflects the fact that the creation of a hole during optical
absorption abruptly lowers the \electron-level.

To relate ${\cal G}_{\electron \electron}^\sigma (\nu)$ to 
a similarly-defined correlator 
${\cal G}_{\FermiR \FermiR}^\sigma (\nu)$, we note that 
the anomalous time-dependence of $c_{k \sigma} (t)$
implies the equation of motion
\begin{eqnarray}
\label{eq:EOMoperator}  
i \partial_t  c_{k \sigma}(t) & = & \mbox{[}
c_{k \sigma}(t), H^{\initial} \mbox{]} 
+ c_{k \sigma}(t) (\varepsilon_{\hole \bar \sigma} - \Ueh n_\electron) \; , 
\\ 
& = & c_{k \sigma} (t) [ \varepsilon_{k \sigma} + \varepsilon_{\hole \bar \sigma}
- \Ueh n_\electron ] +  v e_\sigma (t) \; ,  \qqph 
\end{eqnarray}
where $v = \sqrt {\Gamma / \pi \rho}$. Inserting this into the
definition of $G^\sigma_{kk'}(t)$ one finds
\begin{eqnarray}
\label{eq:EOM1}
 i \partial_t G^\sigma_{kk'}(t) \sim 
[ \varepsilon_{k \sigma} + \varepsilon_{\hole \bar \sigma}]
G^\sigma_{kk'}(t) + v G^\sigma_{\electron k'}(t) \; , 
\end{eqnarray}
where terms that become subleading for $t \to \infty$ have been
dropped (a term containing $\delta (t)$; and one containing
$\langle c_{k \sigma} (t) n_\electron c_{k' \sigma}^\dag
\rangle_\initial$ which contains more operators and hence decays more
quickly with time than the correlators that were retained).
Similarly one finds 
\begin{eqnarray}
\label{eq:EOM2}
 i \partial_t G^\sigma_{\electron k'}(t) \sim 
[ \varepsilon_{k' \sigma} + \varepsilon_{\hole \bar \sigma}]
G^\sigma_{\electron k'}(t) + v G^\sigma_{\electron \electron }(t) \; , 
\end{eqnarray}
where the cyclic property of the trace was used to write
\begin{eqnarray}
\label{eq:cyclictrace}
 \langle [c_{\electron \sigma }(t) , H^\initial] c_{k'\sigma}^\dag  \rangle_\initial
 & = &
 \langle c_{\electron \sigma} (t) [ H^\initial,  c_{k'\sigma}^\dag ] \rangle_\initial \; . 
\end{eqnarray}
Fourier-transforming Eqs.~(\ref{eq:EOM2}) and (\ref{eq:EOM1}) using the
convention stated just before (\ref{eq:GeetSOM}) and eliminating
${\cal G}_{\electron k'}^\sigma (\nu)$, we readily find the asymptotic
relation:
\begin{eqnarray}
{\cal G}_{k k'}^\sigma (\nu) \sim 
\frac{v^2 {\cal G}^\sigma_{\electron \electron} (\nu)}{
(\nu_+ + \Delta - 
\varepsilon_{k \sigma})
(\nu_+ + \Delta - 
\varepsilon_{k' \sigma}) } \; , 
 \label{eq:Gkk'Gee}
\end{eqnarray}
where $\Delta = \omega_\threshold - \varepsilon_{\hole \bar \sigma} $. 
It follows that
\begin{eqnarray}
\label{eq:GeeGcc}
 {\cal G}^\sigma_{\FermiR \FermiR} (\nu) = \sum_{kk'} {\cal G}_{k k'}^\sigma (\nu) \sim
- \pi \rho \Gamma {\cal G}_{\electron \electron}^\sigma (\nu) \; , 
\end{eqnarray}
where the double sum $\sum_{kk'} = \rho^2 \int {\rm d} \epsilon_{k}
{\rm d} \epsilon_{k'}$ is exhausted by two $\delta$-functions,
since $|\Delta|$ is of order $|\varepsilon^\final_{\electron \sigma}|$,
which we assume to be smaller than the bandwidth $D$.  
Eq.~(\ref{eq:GeeGcc}) implies
\begin{eqnarray}
\label{eq:AGcc}
A^\SC_\sigma (\nu) \sim  \frac{2}{\pi \rho \Gamma}
 {\rm Im} {\cal G}_{\FermiR \FermiR}^\sigma (\nu)  \;  .
\end{eqnarray}
To calculate $G^\sigma_{\FermiR \FermiR}(t)$ for $t \gg 1/\Tk$, we may
now replace $H^\initial \to H_{\FermiR}$ and $H^\final \to
H^\ast_\SC + H^\prime_\SC$ in Eq.~(\ref{eq:AGcc}):
\begin{equation}
  G_{\FermiR \FermiR}^{\sigma}(t)
\sim 
{}_\initial \langle \ground | \ex{iH_\FermiR t}
c_{\sigma} \ex{-iH_{\SC}^\ast t}
c_{\sigma}^\dagger
| \ground \rangle_\initial
\; .
\label{eq:respbare1}
\end{equation}
This response function is similar to that calculated in the X-ray edge
problem\cite{Mahan1967, Nozieres1969, Ambrumenil2005}: there absorption of an
X-ray photon excites an atomic core electron into the conduction band
(described by $c^\dagger_\sigma$), leaving behind a core hole which
constitutes a scattering potential with respect to $H_{\FermiR}$
(described by $H_{\SC}^\ast$). The calculation of
Eq.~(\ref{eq:respbare1}) is standard (e.g. {\Ambrumenil}) and yields (we
show only the leading power law)
\begin{eqnarray}
\label{eq:Gcct-result}
  G^\sigma_{\FermiR \FermiR} (t) & \sim &
 t^{-[(\delta_\sigma - \pi)^2 + \delta^2_{\bar \sigma}]/\pi^2}
  \; . \qphan
\end{eqnarray}
where $\delta_\sigma = \delta_\sigma (0)$ denotes the 
phase shifts at the Fermi energy.
 This power-law has an instructive interpretation, due
  to Hopfield{\Hopfield}: according to Anderson's
  orthogonality catastrophe{\Anderson}, two Fermi seas
  subject to different local scattering potentials that cause their
  modes to differ in phase by $\delta_\sigma (\varepsilon_{k\sigma})$,
  have a ground state overlap ${}_\final\langle G | G \rangle_\initial
  \sim L^{- \sum_\sigma \delta^2_{\sigma}(0)/ \pi^2}$ which vanishes
  in the limit of system size $L \to \infty$. In analogy,
  Eq.~(\ref{eq:Gcct-result}) can be viewed as the overlap ${}_\final \langle
  G | c^\dag_\sigma |G\rangle_\initial$ for systems of size $L \propto
   t$: the effect of $c^\dag_\sigma$, which puts an extra spin-$\sigma$
  electron at the scattering site at $t=0$, is analogous to having an
  additional infinitely strong scatterer of $\sigma$-electons in the
  initial, but not the final state, implying an extra shift $-\pi$ for
  the phase difference of the $\delta_{\sigma}$ modes.

\begin{figure}
\centering
\includegraphics[width=0.8\linewidth]{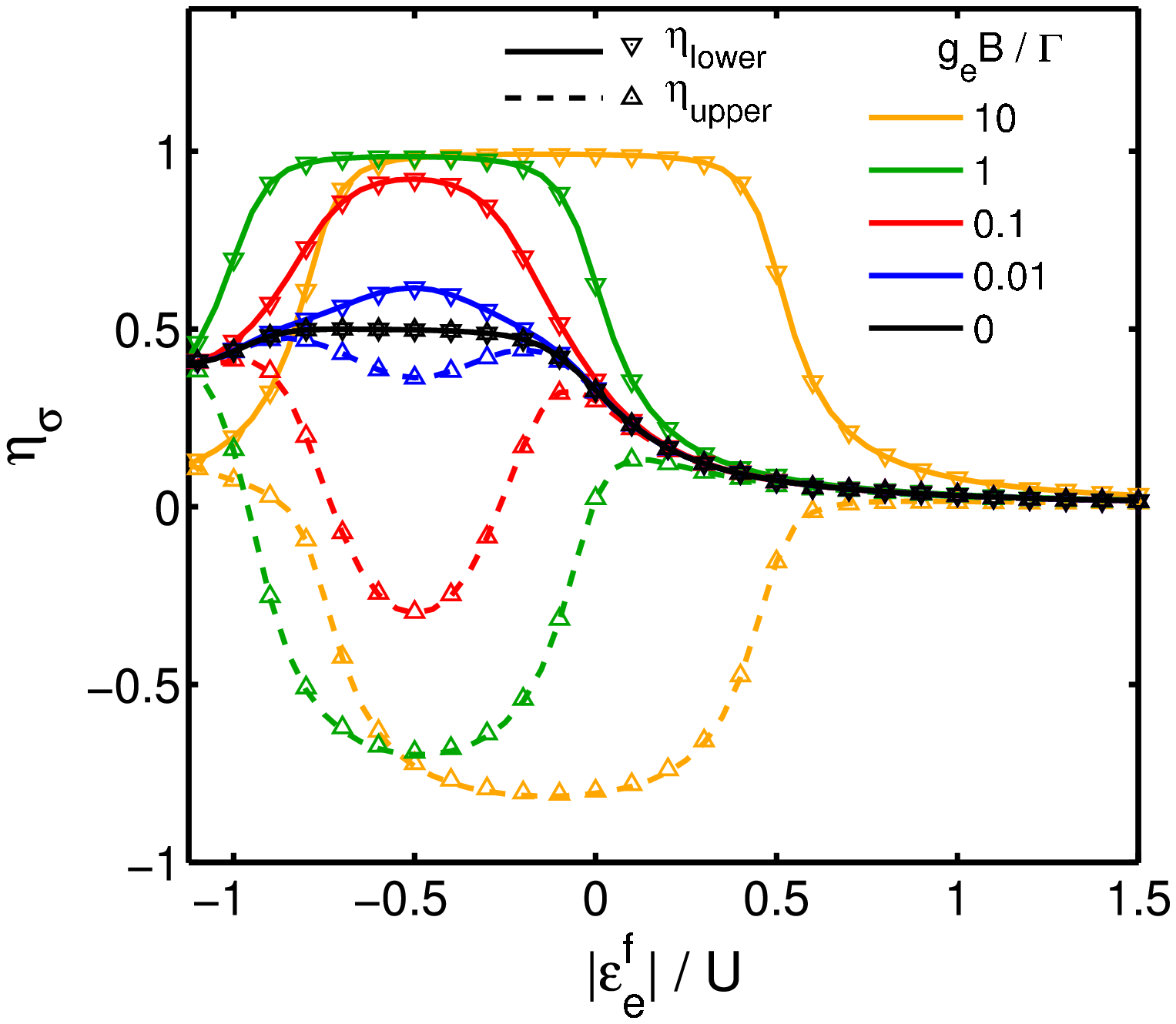}
\caption{{\bf Checking Eq.~({\eqHopfield}), Hopfield's rule of thumb for the singularity
  exponent $\eta_{\sigma}$ of the Kondo-exciton, for various
  combinations of $|\varepsilon^\final_\electron|$ and $B$}.  Solid
  lines were calculated via the the r.h.s. of Eq.~({\eqHopfield}), using
  NRG-results for $\Delta n_{\electron \sigma}$ as input; symbols give
  the values of the exponent $\eta_\sigma$ extracted from $A_\sigma
  (\nu \to 0)$. Symbols agree with solid lines to within 1~\%,
  confirming the validity of Hopfield's rule of thumb for absorption
  into a Kondo-correlated final state. NRG parameters: $\Uee=0.1D$, $\Gamma=0.1 \Uee$, $T= 0$, $\Lambda = 2.3$, Kept states: $1700$, $\alpha = 0.6$.}
\label{fig:HopfieldSI}
\end{figure}

The phase shifts at the Fermi energy,
needed in Eq.~(\ref{eq:Gcct-result}), are given by $\delta_\sigma  =
\pi \Delta n_{\electron \sigma} $, according to the Friedel sum rule\cite{Friedel1956, Langreth1966}, valid for $T=0$ and for arbitrary values
of $B$, $\bar n^\final_\electron $ and $\bar n^\initial_\electron
$. Collecting results, we find from Eqs.~(\ref{eq:AGcc}) and
(\ref{eq:Gcct-result}) that
\begin{equation}
\label{eq:ASCsi}
A^\SC_\sigma (\nu) \sim \Tk^{-1} (\nu / \Tk)^{- \eta_\sigma} \; ,
\end{equation}
with the infrared singularity exponent $\eta_\sigma$ given by
Eq.~({\eqHopfield}), 
\begin{equation}
\label{eq:exponentsmalldetuning-SOM}
\eta_\sigma =
2 \Delta n_{\electron \sigma} - \sum_{\sigma'} (\Delta n_{\electron \sigma'})^2
 \; .
\end{equation}
The dimensionful prefactor in \Eq{eq:ASCsi} was adjusted to ensure that
Eqs.~(\ref{eq:ASC}) and (\ref{eq:A-LM-results}) match, up to numerical prefactors, at the
crossover scale $\nu = \Tk$, implying a prefactor $(\rho \Gamma
)T_{\rm K}^{\eta_{\sigma} - 1}$ in Eq.~(\ref{eq:Gcct-result}).
We have checked Eq.~({\eqHopfield}) numerically for a range of parameter
combinations, see Fig.~5b and Fig.~\ref{fig:HopfieldSI}, finding it to
hold to within 1~\%.

\section*{Appendix 7: Magnetic field dependence of the absorption lineshape}

In the main text, we pointed out that as $|B|$ increases past $\Tk$,
the evolution of the 
infrared exponents $\eta_\sigma$ with $B$ depends
on whether the spin $\sigma$ of the photoexcited electron
matches that of the lower or upper of the Zeeman-split e-levels,
which we distinghuished by writing $\sigma =$~lower  or
upper, respectively. Here we discuss this $\sigma$-dependence
in some more detail.

 It is instructive to 
express $\eta_\sigma$ of Eq.~({\eqHopfield}) in terms of the final occupation 
$\bar n^\final_\electron$ and 
magnetic moment 
$\magnetization$ of the e-level, 
writing $\bar n^\final_{\electron \sigma } = \frac{1}{2} \bar n^\final_\electron 
+  \sigma m^\final_\electron$: 
\begin{equation}
\eta_\sigma = \eta_0  + 2 \magnetization \sigma
- 2 (\magnetization)^2\; , \quad \eta_0 = \bar n^\final_\electron 
(1 - \frac{1}{2} \bar n^\final_\electron ) \; . 
\label{eq:eta-m-main}
\end{equation}  
In particular, this implies the simple relation $m^\final_\electron =
\frac{1}{4} (\eta_+ - \eta_-)$.  Moroever, at $\bar n^\final_\electron
= 1$, $m^\final_\electron $ is a universal function of $ g_\electron
B/\Tk$; hence, the same is true for the infrared singularity exponents
$\eta_\sigma$.  (At very large fields, however, a bulk Zeeman field,
neglected above, will spoil universality, see Appendix~8.)

For $|B| \ll \Tk$, the magnetic moment is determined by the linear
static susceptibility $\chi_0 = 1 / 4 \Tk$ via $\magnetization = -
g_\electron B \chi_0$, implying $\eta_\sigma = \eta_0 - \frac{\sigma
  g_\electron B}{2 \Tk}$, to lowest order in $B/\Tk$.  As $|B|$
increases past $\Tk$ and the magnetization tends towards
$\magnetization \simeq - \frac{1}{2} {\rm sgn} (g_\electron B)$, the
exponents tend to $\eta_\sigma \to -\frac{1}{2} (\bar
n^\final_\electron - 1)^2 - \sigma \, {\rm sgn}(g_\electron B)$. Thus,
they differ by 2 and \emph{attain opposite signs} (since $\bar
n^\electron_\final \in [0,2]$), depending on whether the photoexcited
electron has spin $\sigma = \mp \, {\rm sgn}(g_\electron B)$ (selectable
by choice of circular polarization of the incident ligth), i.e.\
whether its spin matches that of the lower or upper of the
Zeeman-split e-levels, respectively. (Correspondingly, the
photoexcited electron will be said to have an ``$\lowerlevel$''- or
``$\upperlevel$-level-spin'', with exponent $\eta_{\lowerlevel}$ or
$\eta_\upperlevel$.)  This implies a dramatic $\sigma$-dependence of
the evolution of the lineshape $A_\sigma (\nu ) \propto
\nu^{-\eta_\sigma}$ with increasing $|B|$ (Fig.~5a).  In the main text
we focussed explicitly on the case that $\bar n^\final_\electron = 1$,
for which $\eta_{\lowerlevel/ \upperlevel} $ crosses over from
$\frac{1}{2}$ at $B=0$ to $\pm 1$ for $|B| \gg \Tk$ (Eq.~(9) and Fig.~5b):
correspondingly the near-threshold singularity either becomes
stronger, tending towards $\nu^{-1}$; or it becomes weaker, and once
$\eta_\upperlevel$ turns negative, changes to an increasingly strong
power-law decay, tending toward $\nu^{+1}$. As mentioned in the main
text, the difference between the two cases reflects the fact that in
the limit $|B| \gg \Tk$, Anderson orthogonality is absent or maximal
for $\sigma =$~lower or upper, respectively.

In this limit, the dependence of the lineshape on detuning and on
$\sigma$ may also be interpreted in the following, alternative way, in
terms of detuned transitions into the e-level: \emph{At the absorption
  edge}, the photo-excitation of an upper-level-spin electron can be
viewed as being a virtual excitation to the upper Zeeman level (with
effective detuning given by $\sim (\nu - g_\electron B$)) followed by
a spin-flip of the e-level electron assisted by the creation of a FR
electron-hole pair.  Since the phase-space for creating these
electron-hole pairs scales as $\nu$, the overall absorption rate for
$\sigma = \upperlevel$ scales as $\nu / (\nu - g_\electron B)^{2} \sim
\nu / (g_\electron B)^2$. In contrast, for the photo-excitation of a
lower-level-spin electron, the effective detuning is $\nu$, and the
excess energy is deposited into the Fermi sea via the creation of
spin-preserving particle-hole pairs, for which the phase space again
scales with $\nu$.  This leads to an absorption rate for $\sigma =
\lowerlevel$ that scales as $\nu /\nu^2 = 1/\nu$. This explains the
difference in power laws, $\eta_\lowerlevel - \eta_\upperlevel = 2$.

The same picture works for larger photon energies \emph{above the
  edge}, except that for the upper-level-spin electron the energy
deficit in the virtual state gets smaller. Once the energy deficit
reaches zero, a resonant peak in the absorption results. Its width is
determined by the Korringa relaxation rate for the upper e-level in
the dot. In the scaling regime, the Korringa relaxation rate at large
Zeeman energy $(|g_\electron B| \gg \Tk, T)$ is of the order of
$g_\electron B/ \ln^2(g_\electron B/ \Tk)$. Thus, the smaller the
exchange interaction with the band (i.e.\ the smaller $\Tk$), the
sharper the upper-level-spin absorption peak, and the smaller the
weight outside the peak, near the absorption edge.

\section*{Appendix 8: Justification of neglecting the bulk magnetic field}

\begin{figure}
\centering
\includegraphics[width=0.8\linewidth]{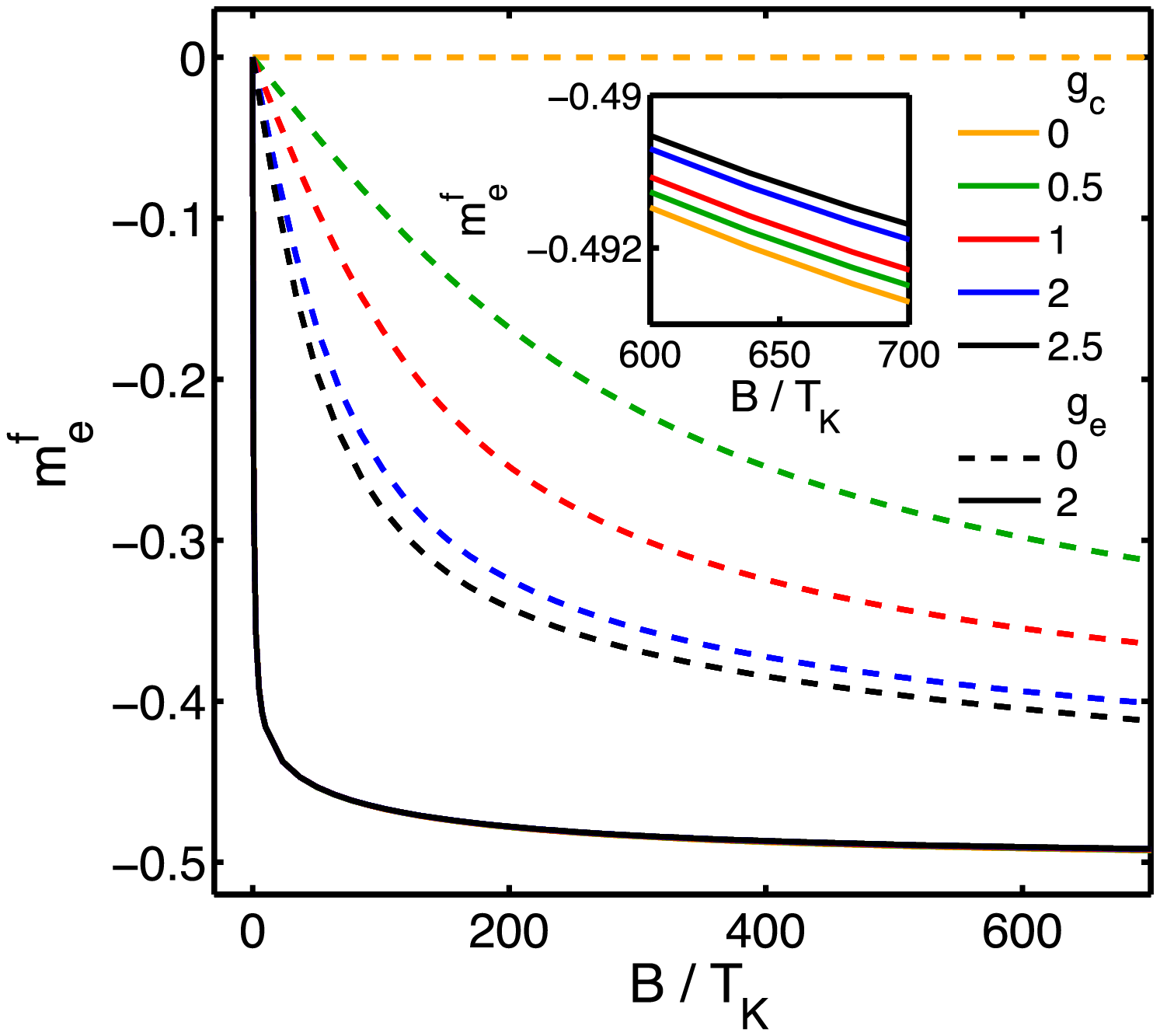}
\caption{{\bf Dependence of magnetization on the bulk $g$-factor
  $g_\FermiR$}.  The final magnetization $m^\final_\electron$ is
  plotted as function of $ B/\Tk$ for several different value of
  $g_\FermiR$ and two choices of $g_\electron$: if $g_\electron = 0 $
  (dashed lines), the magnetization is sensitive to $g_\FermiR$, but
  the crossover scale beyond which it approaches saturation is
  much larger than $\Tk$.  If $g_\electron = 2$ (solid lines), we
  obtain the standard magnetization curve (also shown in Fig.~5b,
  dash-dotted line), for which the crossover scale is $\Tk$. It's
  dependence on $g_\FermiR$ is so weak that it is not discernable
  unless magnified (see inset). NRG parameters: $\Uee=0.1D$, $\varepsilon^\initial_{\electron} = 0.75U$, $\varepsilon^\final_{\electron} = -0.5 \Uee$, $\Gamma=0.1 \Uee$, $\Tk = 3.9 \cdot 10^{-2} \, \Gamma$, $T= 0$, $\Lambda = 2.3$, Kept states: $1024$.}
\label{fig:roleofgc}
\end{figure}

A magnetic field applied parallel to the Fermi reservoir causes a
Zeeman shift not only for the e- and h-levels, but also for the
c-electrons in the Fermi reservoir, with $g$-factors of comparable
magnitude: $g_\electron \approx - 0.6$-0.7, $g_\hole \approx 1.1$-1.2
and $g_\FermiR \approx - 0.2$-0.4. Nevertheless, the effect of a bulk
field, described by $g_\FermiR B \sum_{k \sigma} \frac{1}{2} \sigma
c^\dagger_{k \sigma} c_{k \sigma}$, was neglected in the main text.
The justification for this is as follows{\Garst}: the effect
of a bulk field on Kondo physics is to polarize the Fermi reservoir,
causing the quantum dot to interact with a nonzero net spin $\langle
s_\FermiR^z \rangle \approx - \rho g_\FermiR B/2$.  As a result, the
effective Kondo coupling $\frac {J}{\rho} \vec s_\FermiR \cdot \vec
s_\electron$ generates an extra contribution $- (J g_\FermiR/2) B
s_\electron^z $ to the local spin Hamiltonian, implying an effective
local $g$-factor of
\begin{equation}
  \label{eq:effective-g-factor}
  g^\eff_\electron = g_\electron - g_\FermiR J / 2 \; .  
\end{equation}
(We have checked that a similar result is obtained within the framework
of the single-impurity Anderson model by using the scaling approach of
Haldane to calculate the energy difference between having the local
level occupied by a spin $+$ or $-$.)  Since the
second term in Eq.~(\ref{eq:effective-g-factor}) involves the
\emph{bare}, not the renormalized exchange constant, its effect is
small. 

A more detailed scaling analysis for $|B| \gg \Tk, \nu$ leads
to the conclusion{\Garst} that for large magnetic fields the
effective $g$-factor has the form
\begin{equation}
  \label{eq:scaled-effective-g-factor}
  g_\electron^\eff (B) = g_\electron \left[1 - \frac{1}{2 \ln (B / \Tk)} \right ]
+ (g_\electron - g_\FermiR) J / 2 \; .  
\end{equation}
The factor in square brackets describes a reduction of
$g^\eff_\electron $ with decreasing field that reflects the onset of
the screening correlations. In the so-called scaling limit where the
coupling $J$ is sent to zero and the bandwidth to infinity in such a
way that $\Tk$ remains fixed, the second term becomes negligible for
all fields smaller than the crossover field $B^\ast \approx \Tk
e^{1/J}$.  As a consequence, the local moment $m_\electron$, calculated
as a function of $B$, is expected to be \emph{independent} of
$g_\FermiR$ for all $|B| \lesssim B^\ast$; in other words,
$m_\electron$ will be a universal function of $g_\electron B$ 
in this regime.

We have checked this expectation within the framework of a
single-impurity Anderson model using NRG, proceeding as follows: The
effect of a bulk field is to shift the Fermi reservoirs by $-
\frac{1}{2} \sigma g_\FermiR B$ with respect to the Fermi energy at
$\varepsilon_\Fermi = 0$.  We thus used shifted band edges,
$D^{\max/\min}_\sigma = \pm D - \frac{1}{2} \sigma g_\FermiR B$, and
employed a $\sigma$-dependent Wilsonian logarithmic energy grid for
each of the four energy intervals $[0, D^{\max}_\sigma]$ and
$[D^{\min}_\sigma,0]$ (with $\sigma = \pm$). For the resulting Wilson
chain we calculated the local moment $m_\electron (B)$ for both
$g_\electron = 0 $ and 2, for several different choices of
$g_\FermiR$, see Fig.~\ref{fig:roleofgc}.  As expected, the effect of turning on 
$g_\FermiR $, though noticable for large fields if $g_\electron = 0$, 
is very small if $g_\electron = 2$.

\begin{figure}
\centering
\includegraphics[width=0.8\linewidth]{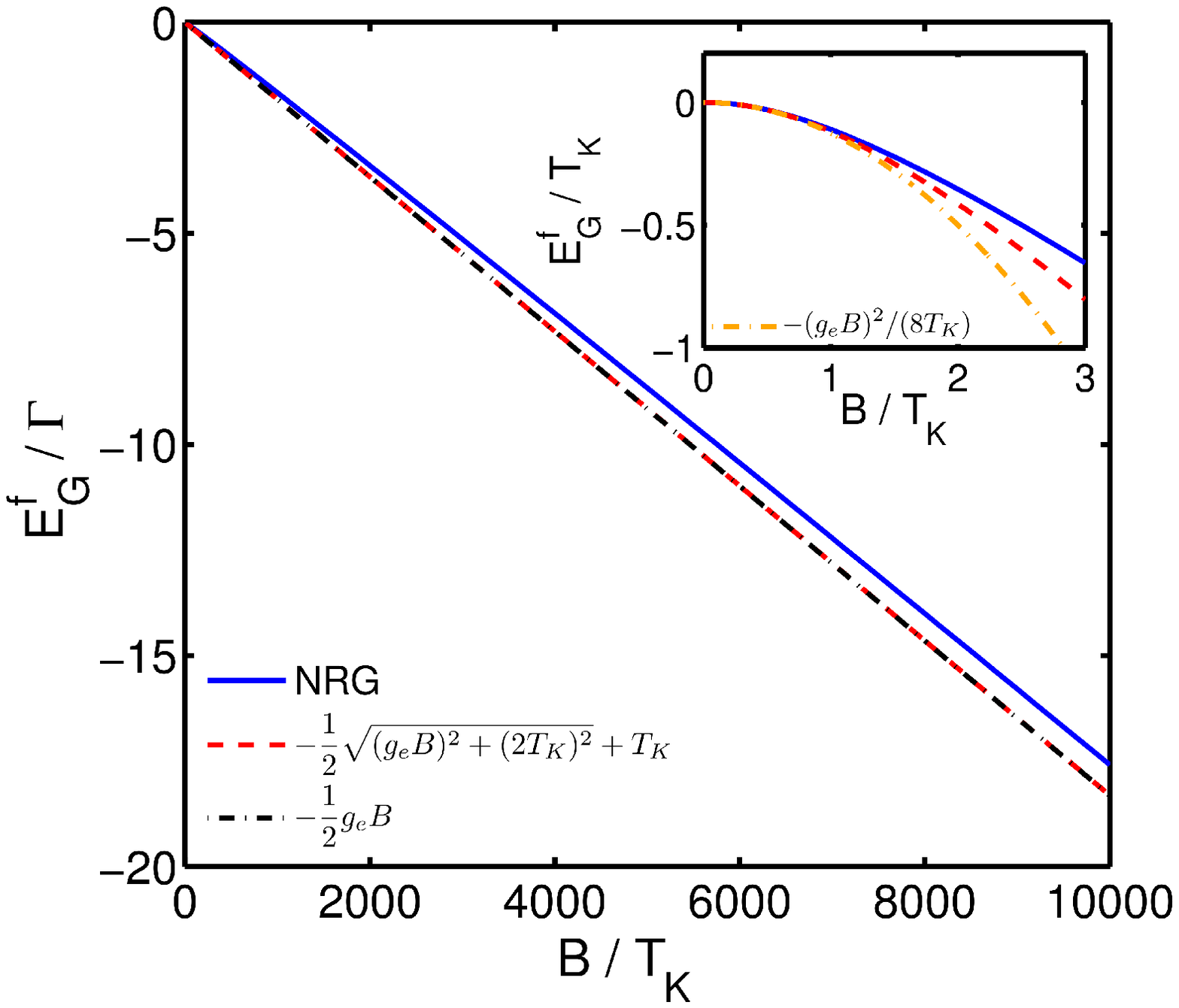}
\caption{{\bf Dependence of the ground state energy on magnetic field}. The
  final-state ground state energy (ground state energy of
  $H^{\final}$), $E_{G}^{\final}$, is plotted as a function of $B/\Tk$
  as calculated by NRG (blue) and compared to the analytic expressions
  given by Eq.~(10) (red). Inset magnifies the region $B \ll \Tk$ to
  show the quadratic dependence of the ground state energy (and hence
  the threshold) on the magnetic field. At large magnetic fields, the slope of the NRG curve differs slightly from that of $-\frac{1}{2} g_\electron B$ due to slight ($B$-dependent) renormalizations of the e$\sigma$ level positions caused by its
by hybridization with the FR. NRG parameters: $\Uee=0.1D$, $\varepsilon^\initial_{\electron} = 0.75U$, $\varepsilon^\final_{\electron} = -0.5 \Uee$, $\Gamma=0.062 \Uee$, $\Tk = 3.7 \cdot 10^{-3} \, \Gamma$, $T= 0$, $\Lambda = 2.3$, Kept states: $1024$.}
\label{fig:checkishii}
\end{figure}

\end{appendix}

\bibliographystyle{naturemag}

\begin{thebibliography}{18}
\bibitem{foot1} This is naturally true only for time-scales longer than the inverse bandwidth of the fermionic reservoir.
\bibitem{foot2} The pre-factor in $A^\SC_\sigma (\nu)$ can be understood as a consequence of the
  conservation of QD transition oscillator strength.
\end{thebibliography}

\end{document}